\documentclass[letterpaper,twocolumn,10pt]{article}
\usepackage{usenix}
\usepackage{booktabs} %
\usepackage[inline]{enumitem}
\usepackage{url}
\usepackage{comment}
\usepackage[font=small,labelfont=bf]{caption}
\usepackage{subcaption}
\usepackage{graphicx}
\usepackage[scaled]{beramono}
\usepackage{ifthen}
\usepackage{amsmath, amssymb}
\usepackage{tikz}
\usepackage{xspace}
\usepackage{verbatim}
\usepackage{booktabs}
\usepackage{hyperref}
\usepackage{float}
\usepackage{multirow}
\usepackage{soul}
\usepackage{mathtools}
\DeclarePairedDelimiter\floor{\lfloor}{\rfloor}

\newcommand{\alglinelabel}{%
  \addtocounter{ALC@line}{-1}%
  \refstepcounter{ALC@line}%
  \label%
}

\hypersetup{
  colorlinks=true,      %
  linkcolor=blue,       %
  citecolor=magenta,    %
  filecolor=cyan,       %
  urlcolor=red          %
}
\newenvironment{denseitemize}{
\begin{itemize}[topsep=2.5pt, partopsep=0pt, leftmargin=1.5em]
  \setlength{\itemsep}{2.5pt}
  \setlength{\parskip}{0pt}
  \setlength{\parsep}{0pt}
}{\end{itemize}}

\newenvironment{denseenum}{
\begin{enumerate}[topsep=2.5pt, partopsep=0pt, leftmargin=1.5em]
  \setlength{\itemsep}{2.5pt}
  \setlength{\parskip}{0pt}
  \setlength{\parsep}{0pt}
}{\end{enumerate}}

\AtBeginDocument{%
  \providecommand\BibTeX{{%
    \normalfont B\kern-0.5em{\scshape i\kern-0.25em b}\kern-0.8em\TeX}}}

\newcommand{\autops}{{AutoPS}}
\newcommand{\pservice}{{Parameter Service}}
\newcommand{\psmaster}{{pMaster}}
\newcommand{\pserver}{{Aggregator}}
\newcommand{\psagent}{{Agent}}

\newcommand{\lowperf}{\texttt{LowPerf}}

\def\ie{{i.e.}}
\def\eg{{e.g.}}

\def\mxnet{{MXN}et}
\def\tensorflow{{T}ensor{F}low}

\def\vgg#1{VGG$#1$}
\def\alexnet{AlexNet}
\def\resnet#1{ResNet$#1$}

\def\lm{AWD-LM}
\def\bert{BERT}

\def\imagenet{ImageNet}
\def\wikidata{WikiText-2}
\def\bertdata{BookCorpus}

\newboolean{showcomments}
\setboolean{showcomments}{true}
\newcommand{\mosharaf}[1]{  \ifthenelse{\boolean{showcomments}}
{ \textcolor{red}{(Mosharaf:  #1)}} {}  }
\newcommand{\juncheng}[1]{  \ifthenelse{\boolean{showcomments}}
{ \textcolor{blue}{(Juncheng:  #1)}} {}  }
\newcommand{\kang}[1]{  \ifthenelse{\boolean{showcomments}}
{ \textcolor{purple}{(Kang:  #1)}} {}  }

\begin{document}
\title{Elastic Model Aggregation with Parameter Service}
\author{
{\rm Juncheng Gu$^1$, Mosharaf Chowdhury$^1$, Kang G. Shin$^1$, Aditya Akella$^2$}\\
\and
$^1$University of Michigan, $^2$University of Texas at Austin
} %

\maketitle
\pagestyle{plain}
\begin{abstract}
Model aggregation -- the process that updates model parameters -- is an 
important step for model convergence in distributed deep learning (DDL).
However, the parameter server (PS), a popular paradigm of performing 
model aggregation, causes CPU underutilization in deep learning (DL) 
clusters, due to the bursty nature of aggregation and static resource allocation.
To remedy this problem, we propose \emph{Parameter Service}, an elastic model 
aggregation framework for DDL training, which decouples the function of model 
aggregation from individual training jobs and provides a shared model 
aggregation service to all jobs in the cluster.
In {\pservice}, model aggregations are efficiently packed and dynamically 
migrated to fit into the available CPUs with negligible time overhead.
Furthermore, {\pservice} can elastically manage its CPU resources based on 
its load to enhance resource efficiency.
We have implemented {\pservice} in a prototype system called {\autops} and evaluated 
it via testbed experimentation and trace-driven simulations.
{\autops} reduces up to $75\%$ of CPU consumption with little or no performance 
impact on the training jobs.
The design of {\pservice} is transparent to the users and can be incorporated 
in popular DL frameworks.

\end{abstract}

\section{Introduction}
\label{a_sec:intro}
\vspace{-5pt}
Deep learning (DL) has become a major driving force in many application 
domains, including image classification, speech recognition, and machine 
translation~\cite{resnet, hinton2012deep, bert-google}.
With increasing model complexity and large training datasets, many DL models are trained distributedly.
Distributed deep learning (DDL) is 
becoming popular~\cite{easeml-18, atc19-shared-cluster} to meet their ever-growing demands.

Although GPU scheduling has so far received the most attention in the context 
of DL clusters~\cite{themis-nsdi20, tiresias-nsdi19, gandiva, optimus} and 
rightly so, we observe that the impact of traditional cluster resources such as CPU can be significant (\S\ref{a_sec:motivation}).
This is especially true for DDL training jobs, where parameter servers run 
on traditional virtual machines (VMs) or containers.
When running those jobs, up to $80$\% of the allocated CPUs may be wasted 
due to the bursty and periodic nature of model aggregation.
Note that CPUs are not free to use;
renting one CPU core in the cloud takes around $\$900$/yr~\cite{cpu-cost-nsdi18}.
Given the increasing number of DDL training jobs, 
users spend a substantial amount of money on the rented-but-unused CPUs.

The root cause of this CPU under-utilization is the "blind" application of resource management 
techniques from big data clusters for emerging DDL jobs.
A VM or container allocated to a parameter server has a \emph{fixed} amount 
of CPU resource, which is exclusive (non-sharable) and provisioned for 
peak usage. 
However, a parameter server usually cannot always keep its assigned CPU fully utilized -- CPU consumption by DDL training is inherently bursty.
Parameter servers of a DDL job must idly wait for the model updates from workers which 
are generated layer by layer according to the progress of 
back-propagation. 

To remedy this inefficiency, we propose {\pservice}, an elastic model aggregation 
framework for DDL training that aims to improve overall CPU utilization 
without sacrificing training performance. 
Unlike prior work on model aggregation~\cite{osdi14ps, tensorflow-osdi16} 
where parameter servers are assigned to individual jobs, 
{\pservice} \emph{decouples} model aggregation from training and exposes a shared 
model aggregation service to all training jobs for better CPU utilization.
{\pservice} resides between the DL framework and the low-level 
infrastructure. Therefore, it is transparent to the users %
and only requires a few modifications to the DL framework.

\begin{figure*}[!t]
  \begin{subfigure}[t]{0.4\textwidth}
      \centering
      \includegraphics[scale=0.5]{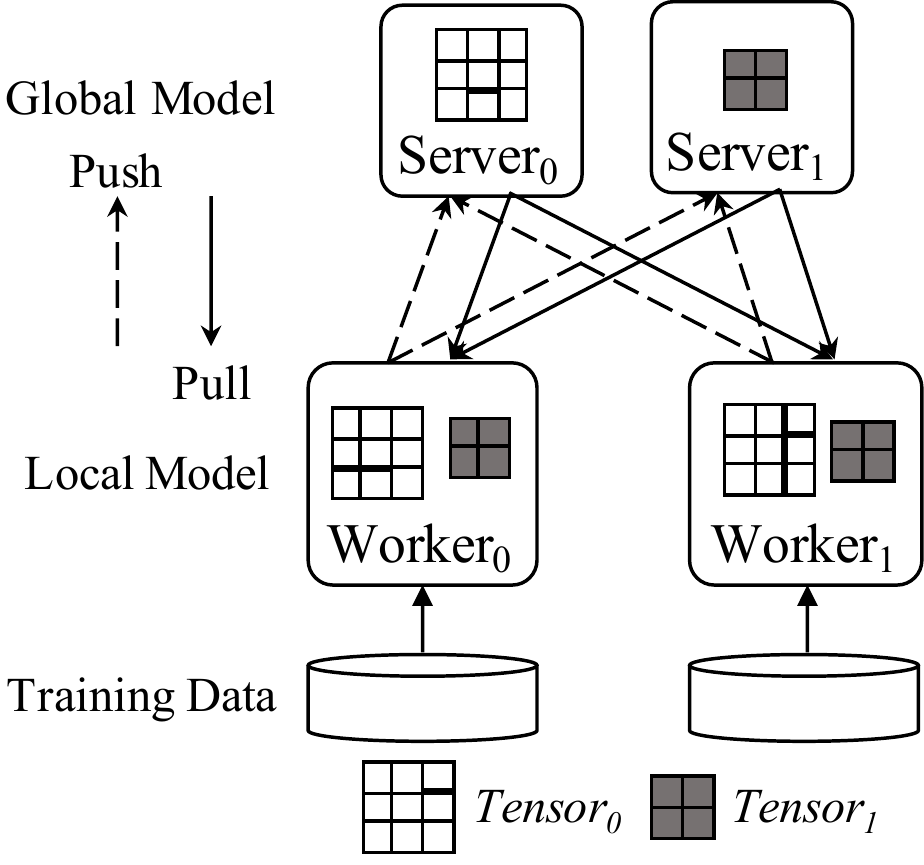}
      \caption{Job components.}
      \label{a_fig:toy-ps}
  \end{subfigure}
  \begin{subfigure}[t]{0.55\textwidth}
      \centering
      \includegraphics[scale=0.5]{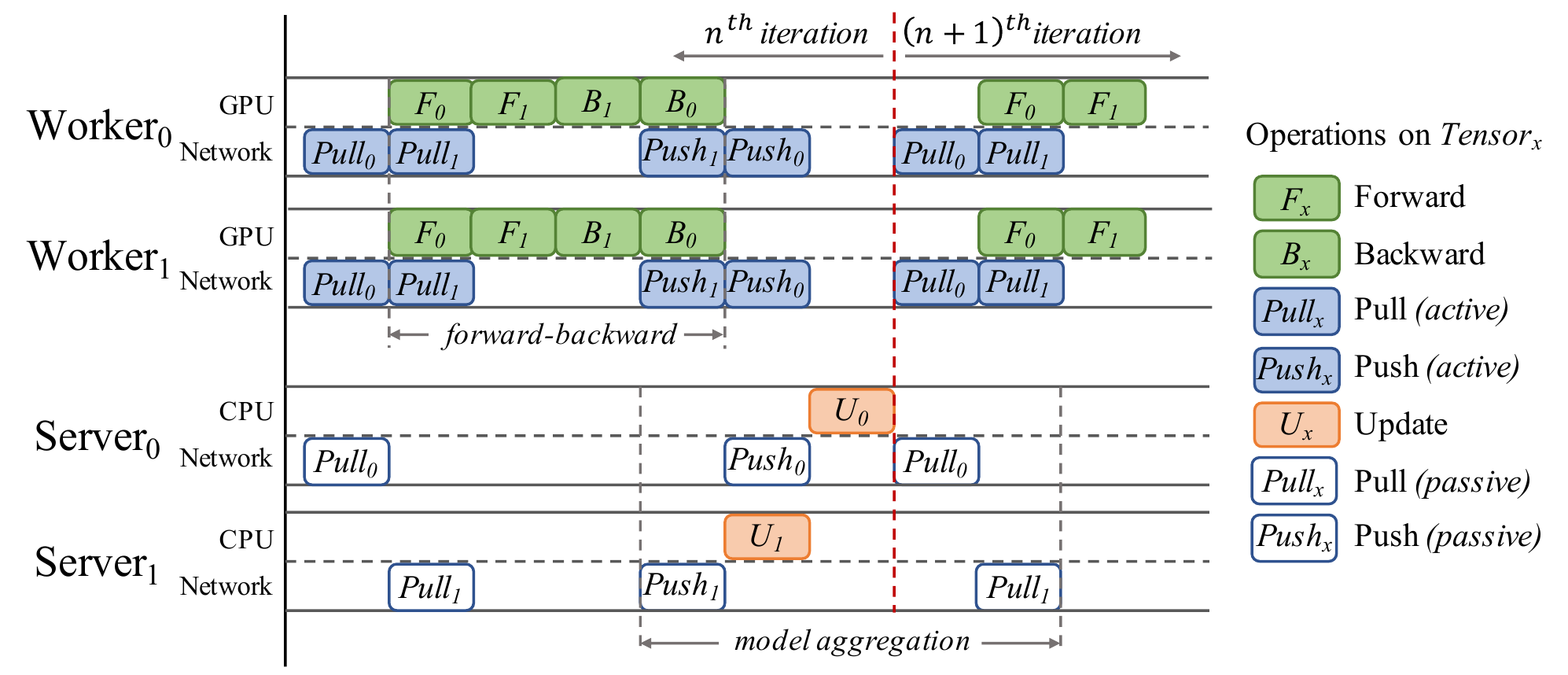}
      \caption{Execution timeline.}
      \label{a_fig:toy-execution}
  \end{subfigure}%
  \caption[A toy example of data-parallel DDL training.]{A toy example of DDL training. There are 2 tensors in this model.
  Data-parallelism and parameter server (PS) for distributed training are applied here.
  There are 2 workers and 2 server instances. 
  Each server instance is in charge of one tensor and handles the related aggregation operations.
  \protect\subref{a_fig:toy-ps} presents the key job components in this example.
  \protect\subref{a_fig:toy-execution} shows the execution timeline of all participants%
  on different resource (\eg, GPU, CPU, and network).
  After the $n^{th}$ iteration, the job proceeds to the $(n+1)^{th}$ iteration with the same operations.
  For simplicity, the width of each operation block is equalized and does not represent the actual execution time.}
  \label{a_fig:toy-example}
\end{figure*}

The crux of the problem is answering: \emph{How to efficiently share CPU resources for model aggregation among DDL 
jobs without degrading their performance?} 
{\pservice} relies on two key knobs to address this question: \emph{dynamic 
workload assignment} and \emph{elastic resource management}. 
It can flexibly pack the model aggregations from the same or different jobs 
onto a single server to fill its idle CPU cycles so as to avoid resource
wastage.  When any workload change occurs (\eg, job arrivals and/or exits), 
{\pservice} can dynamically update the assignments to enhance resource 
efficiency and preserve job performance.
In addition, {\pservice} manages CPU resource elastically at the server level.
The number of servers for model aggregation can be seamlessly scaled 
up or down based on the change of load in {\pservice}.

Moreover, dynamic model aggregation management requires efficient 
orchestration among servers, which would otherwise elongate the execution 
of workers and waste their GPUs.  Leveraging two unique characteristics of DDL 
training, {\pservice} can flexibly re-assign model aggregations with 
negligible time overhead.
Model aggregations happen at the tensor level with no data dependencies 
among each other.  Thus, {\pservice} can freely reassign a single model aggregation 
without interrupting other ones in the same job.
Once a decision is made, the master copy of tensor data 
needs to be migrated from the original server to the new one. 
Because DDL training is done iteratively with a fixed execution pattern, 
{\pservice} hides the time overhead by only performing data migration 
when the job is in the training stage at the worker side.

We propose a heuristic for assignment of model aggregations 
and a simple feedback-based scheme for resource scaling.
Relying on the iterative nature of DDL training, our assignment scheme finds 
proper servers and gets cyclic execution slots for model aggregations.
Incorporating with model aggregation assignment, our resource 
scaling mechanism balances between resource utilization and job performance.
We have implemented {\pservice} in a system called {\autops} and deployed it 
on a real DL cluster. 
We evaluated it using Apache MXNet~\cite{mxnet-arxiv15} with multiple 
state-of-the-art DL models.
{\autops} reduces up to $75\%$ of CPU resource from workload packing, 
with very-limited performance impact on the training jobs.

Overall, we make the following contributions in this paper:
\begin{denseitemize}
    \item {\pservice} decouples model aggregation from training jobs
    and elevates it as a shared cluster-wide service. 
    This makes it easier for users to run DDL training without 
    maintaining parameter servers for each job.
    
    \item {\pservice} resolves the mismatch between DDL training and 
    the infrastructure. 
    It improves CPU utilization by dynamic workload assignment 
    and elastic resource management.
    
    \item {\pservice} is completely transparent to users and requires 
	trivial modifications in DL frameworks.
\end{denseitemize}

\section{Background and Motivation}
\label{a_sec:motivation}
\vspace{-5pt}

In this work, we focus on the classic \emph{data-parallel} DDL training using the parameter server (PS) architecture, 
where multiple workers work on their local copies of the DL model in parallel 
and the training dataset is partitioned across all the workers (Figure~\ref{a_fig:toy-ps}).
The \emph{server} instance in PS is separated from the training workers. 
It hosts the most updated copy (\ie, master copy) of the model parameters,
and updates the model parameters in the master copy using the \emph{aggregated} result.
For performance reasons, a single DDL job may have multiple PS instances,
each of which host different parts (\ie, parameter) of the model (Figure~\ref{a_fig:toy-ps}).
In current PS design~\cite{osdi14ps, bytesheduler}, the assignment of model parameters 
is often \textbf{\emph{static}} once the job starts. 
Thus, the workload on each PS instance will not be changed when the job is running.

\subsection{CPU Underutilization}
\label{a_sec:cpu-issue}
Most DDL training jobs run in shared clusters to attain cost-effectiveness.
Despite its benefits, there is a crucial mismatch between DDL training and 
the resource management software of the infrastructure.
Many CPU cores, that are statically assigned to parameters server containers or VMs,
are wasted due to the bursty model aggregations.

\begin{figure}[!t]
    \centering
    \includegraphics[scale=0.32]{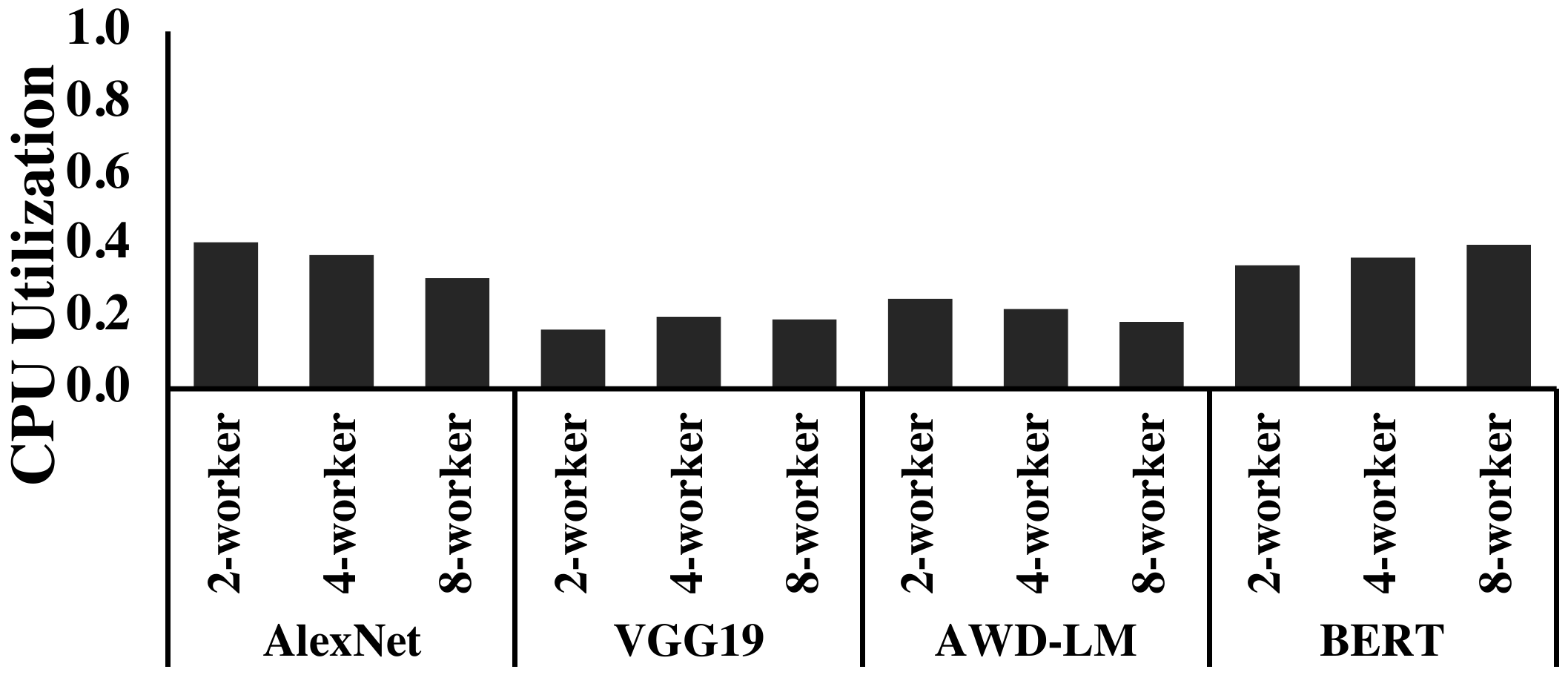}
    \caption[Average CPU utilization of model aggregation.]{Average CPU utilization of model aggregation.
    Each model is trained on {\mxnet} with 1 PS server and different number of workers.}
    \label{a_fig:moti-cpu-util}
\end{figure}

\begin{figure}[!t]
    \centering
    \subfloat[][CPU usage in \vgg19 (\textbf{1}s-2w) training.]{
        \label{a_fig:1s2w-vgg19-cpu}
        \includegraphics[scale=0.2]{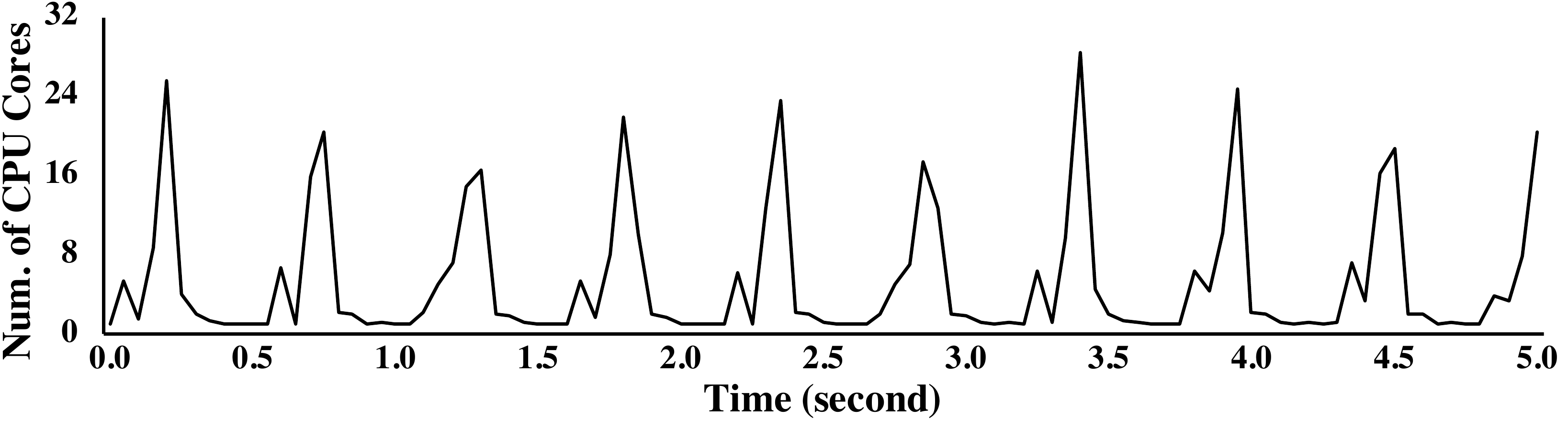}
    }

    \subfloat[][CPU usage in \vgg19 (\textbf{2}s-2w) training.]{
        \label{a_fig:2s2w-vgg19-cpu}
        \includegraphics[scale=0.2]{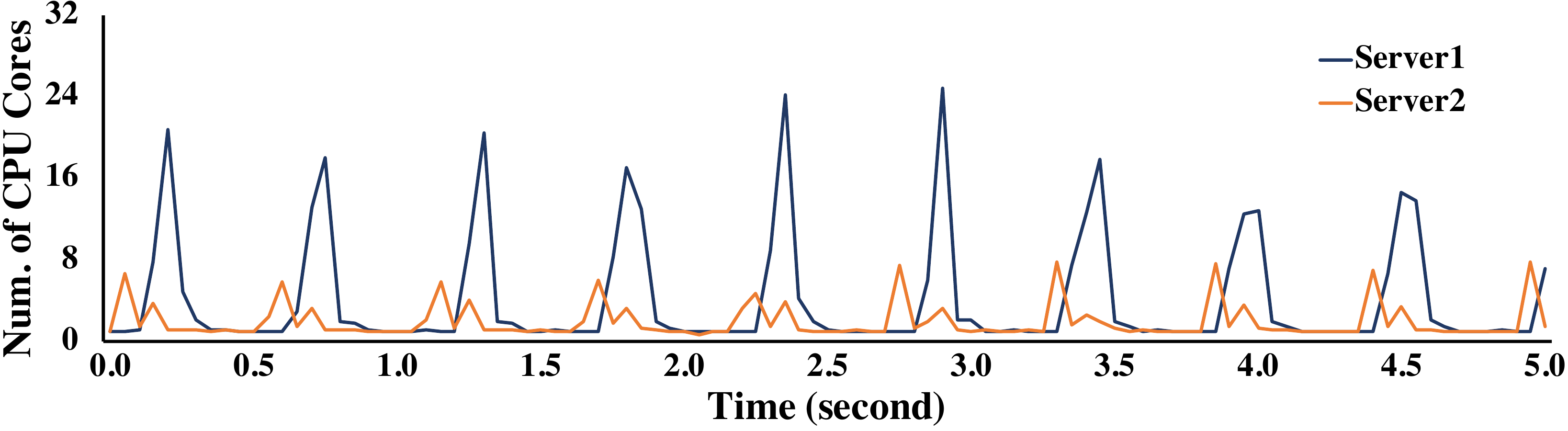}
    }
    \caption[CPU usage of model aggregation in \vgg19 training.]{CPU usage of model aggregation in \vgg19 training.
    \vgg19 is trained on {\mxnet} with 2 different distributed settings: 
    1 PS server and 2 workers, 2 PS servers and 2 workers.
    Servers and workers are distributed to different machines.
    ``1s-2w" means the job has 1 PS server and 2 workers.}
    \label{a_fig:moti-vgg-cpu-trace}
\end{figure}

In each training iteration, the aggregation of a tensor cannot start until 
the completion of backward computation on that tensor.
Due to this dependency, the CPU resource reserved for model aggregation at server 
side is mostly idle when the job is in the \emph{forward-and-backward} stage 
(Figure~\ref{a_fig:toy-execution}).
Besides, model aggregation is performed at tensor-scale. 
There will be a spike of CPU usage when a tensor is ready to be updated.
Therefore, the CPU usage of model aggregation is a combination of sharp spikes 
and idle cycles (Figure~\ref{a_fig:1s2w-vgg19-cpu}).
In a shared cluster, CPU resource are assigned through the abstraction of VM 
or container for isolation.
To achieve good performance, users often oversubscribe the CPU resource
 of their virtual machines (or containers) to satisfy the peak demand.
Due to the mismatch between the dynamic CPU usage and the static CPU allocation, 
CPUs reserved for model aggregation remain underutilized. 

Figure~\ref{a_fig:moti-cpu-util} shows the average CPU utilization of model aggregation 
when training different models.
The number of reserved CPU cores for each job's PS server is the same as 
its peak usage.
Since there is only $1$ PS server in each job, the CPU consumption of model 
aggregation is concentrated at the single PS server.
Among all the jobs, more than a half of the CPU resource are left unused.
For \vgg19 (1s-2w), the average CPU utilization of its server is only $16$\%. 
This issue will get worse when multiple PS servers are used in a single training job.
Since the workload of model aggregation is divided among those PS servers, 
the corresponding CPU consumption is also split (Figure~\ref{a_fig:moti-vgg-cpu-trace}). 
Users need to reserve plenty of CPU cores for each PS server without violating the spikes on each of them.

\vspace{-4pt}
\subsection{Opportunities Brought by DDL training}
\label{a_sec:opportunities}
DDL training has unique characteristics that create opportunities to resolve the mismatch with infrastructure.

\vspace{-10pt}
\paragraph{Tensor-Based Model Structure.} 
Although tensors are layered in the model and have dependencies with 
their neighbors in \emph{forward-and-backward} computation,
they are independent of each other when they are getting aggregated. 
Each tensor can be managed individually at the server side.
Based on this feature, it is possible to apply {\emph{fine-grained and dynamic workload management}}.
Model aggregations can be independently mapped to the proper CPU slots for execution.
The CPU utilization could be improved by packing more model aggregations
from multiple jobs in a single server instance.

\vspace{-10pt}
\paragraph{Iterative Training.}
To handle the aforementioned mismatch, the information of the training job, 
especially model aggregation, is required.
Relying on the iterative feature, the runtime information (e.g., CPU consumption)
of the job measured in the previous iterations can be used as the input 
for making long-term decisions in workload management.

\section{{\pservice}}
\label{a_sec:design}
\vspace{-5pt}
{\pservice} is an elastic model aggregation framework for DDL training.
It aims at enhancing CPU utilization for model aggregation without sacrificing job performance.
In this section, we first present the overview of {\pservice}.
We then illustrate how model aggregations are managed by {\pservice} 
which includes the schemes of workload assignment and resource scaling.

\vspace{-4pt}
\subsection{System Overview}
\label{a_sec:overview}
\vspace{-4pt}
To achieve the aforementioned goals, {\pservice} introduces two features, 
\emph{dynamic workload assignment} and \emph{elastic resource allocation}, to the traditional parameter server-based approach.
When the total load in {\pservice} changes (\eg, job arrival and completion), 
workload reassignment might be triggered if {\pservice} finds any CPU slots 
that fit better for some existing model aggregations.
Meanwhile, incorporating with workload (re)assignment, 
the number of model aggregation servers will be scaled up or down based on demands.

\begin{figure}[!t]
    \centering
    \includegraphics[scale=0.53]{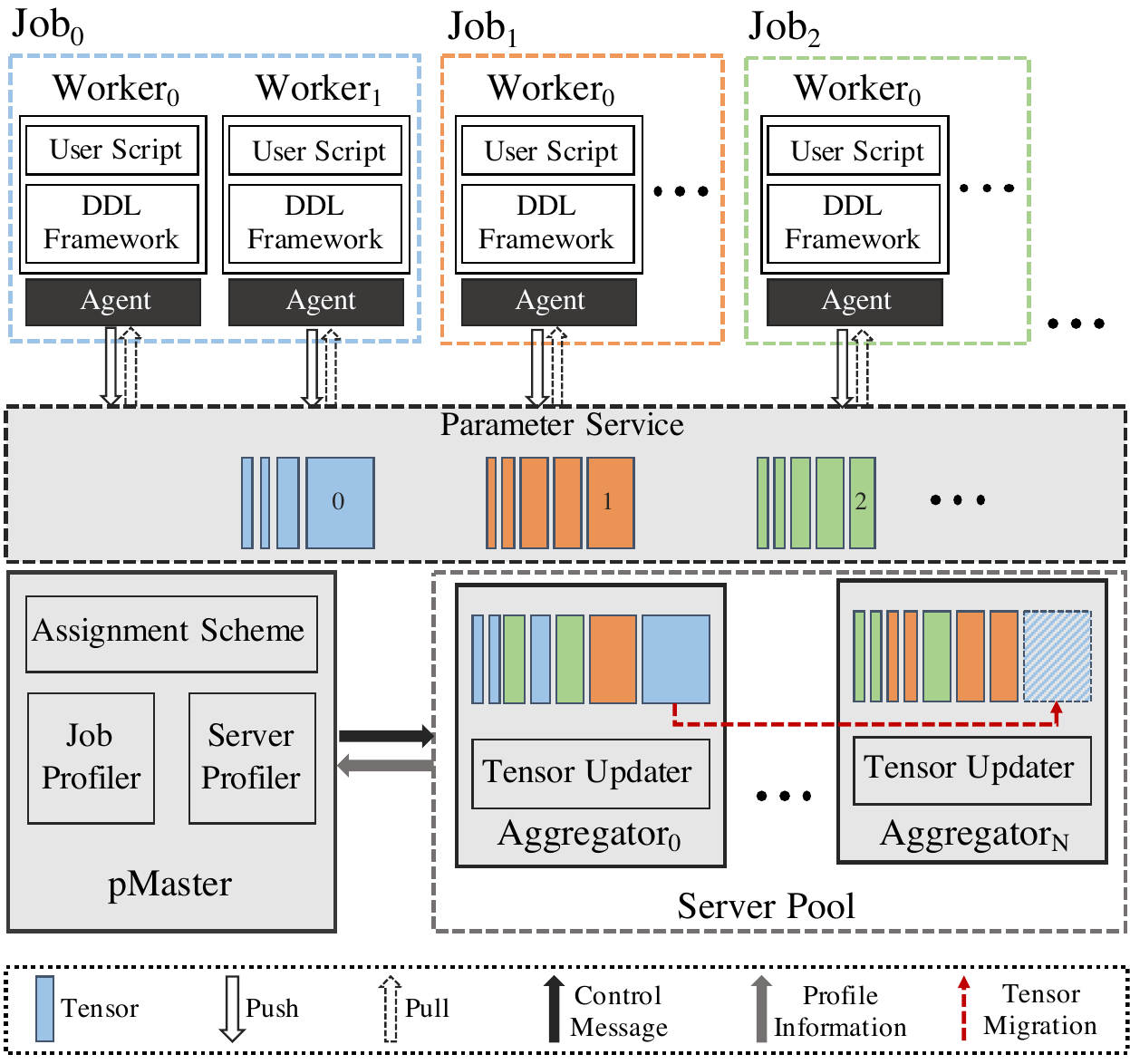}
    \caption[{\pservice} architecture.]{{\pservice} architecture.
    {\psagent} is loaded under the DL framework layer at each worker.
    There is a {\psmaster} in the cluster that manages the pool of {\pserver}.
    Each {\pserver} is placed on an individual machine. 
    }
    \label{a_fig:architecture}
\end{figure}

With {\pservice}, model aggregation is decoupled from individual training jobs,
whereby their workers only need to submit the model aggregation requests to {\pservice} 
through the unique interface and wait for the response, 
without worrying about where the requests are handled and 
how much of resource to be allocated (Figure~\ref{a_fig:architecture}).
In the backend, {\pservice} carefully assigns those requests to the available servers.
In the {\pservice} design, model aggregations are not grouped by training job any
longer; each one is independent and can share the aggregation server with 
the ones from other training jobs.

{\pservice} has three components (Figure~\ref{a_fig:architecture}): 
\begin{denseenum}
  \item \textbf{\emph{{\psmaster}}} is a centralized manager.
  It has a job profiler and a server profiler that keep monitoring the status of training jobs 
  and resource availability (\ie, CPU) of each {\pserver}, respectively.
  When anything is changed, it will adjust the workload assignment and resource allocation accordingly.

  \item \textbf{\emph{{\pserver}}} holds the model tensors of jobs 
  and handles their aggregation requests.
  Any two {\pserver}s can migrate the workload from one to the other, 
  when {\psmaster} reassigns the workload between them.

  \item \textbf{\emph{{\psagent}}} is the interface that {\pservice} exposes to the workers.
  It maintains a table that keeps track of {\pserver}s for tensors in the job.
  When an aggregation request comes, 
  it will forward the request from its worker to the destination by checking the table.

\end{denseenum}

Details on the interactions among those components (\eg, interfaces, and messages) can be found in the appendix.
 
\vspace{-6pt}
\subsection{Tensor Migration}
\vspace{-4pt}
There is one challenge: the aggregation servers keep the master copy 
(latest version) of tensors for model updating (Figure~\ref{a_fig:toy-example}).
Blocked by this data dependency, model aggregations cannot be freely reassigned 
to different servers.

In the current PS systems~\cite{osdi14ps, tensorflow-osdi16}, 
reassigning a single model aggregation task will interrupt the entire training job.
It needs to pause the training process, checkpoint the model parameters, 
and resume the job with the new assignment,
which brings tens of seconds overhead to the training process.
In {\pservice}, we propose a deep-learning-specific migration mechanism that 
migrates tensor data between the {\pserver}s and 
updates assignment information in {\psagent}s.
It exploits the features of DDL training for migration overhead.%

\vspace{-12pt}
\paragraph{Negligible Time Overhead.}
In model aggregation, the master copy of tensor at the {\pserver} side is only needed when it is being updated.
There is a large time window (from the completion of the last \emph{Pull} to the start of \emph{Update}) 
in each iteration that the tensor copy hosted by {\pserver} is not accessed (Figure~\ref{a_fig:toy-execution}).
The actual migration of tensor data is performed within this window to hide its time overhead as much as possible.
In many cases, a tensor migration exposes negligible or even zero time overhead to the training job.
The detailed design of tensor migration protocol can be found in the appendix.

\vspace{-5pt}
\subsection{Model Aggregation Management}
\label{a_sec:algorithm}
\vspace{-4pt}
The core of {\pservice} lies in its schemes of model aggregation assignment and resource scaling that aim at
improving CPU utilization without losing job performance.

When assigning model aggregations, {\pservice} has to carefully balance the trade-off between 
job performance and resource utilization.
Allocating too many {\pserver}s is good for the jobs because of less resource contention but sacrifices resource utilization; and vice versa.
One extreme is the current parameter server solution that allocates individual parameter servers for each training job.
When an {\pserver} has surplus capacity, 
{\pservice} will opportunistically pack more model aggregations on it for resource efficiency. 
It follows the {principle} that a training job should not lose its performance 
when other jobs share {\pserver}s with it.

{\pservice} can elastically change the amount of resource (\ie, the number of {\pserver}s) 
according to the total load of model aggregation in the cluster. 
There are two events that may trigger {\pserver} scaling:
(1) new training job arrival; 
(2) existing job exit.
When a new training job arrives, {\pservice} will assign its model aggregations onto 
the existing {\pserver}s as much as possible.
If they do not fit, new {\pserver}s will be allocated for the extra workload.
For job exit, {\pservice} will return the empty {\pserver}s back to the cluster manager to avoid wastage.
Moreover, it explores the opportunity of freeing the least-loaded {\pserver}s 
by trying to reassign the model aggregations to other {\pserver}s.

\vspace{-4pt}
\subsubsection{Model Aggregation Assignment}
\label{a_sec:assignment}
\vspace{-4pt}

\begin{figure}[!]
  \centering
  \includegraphics[scale=0.35]{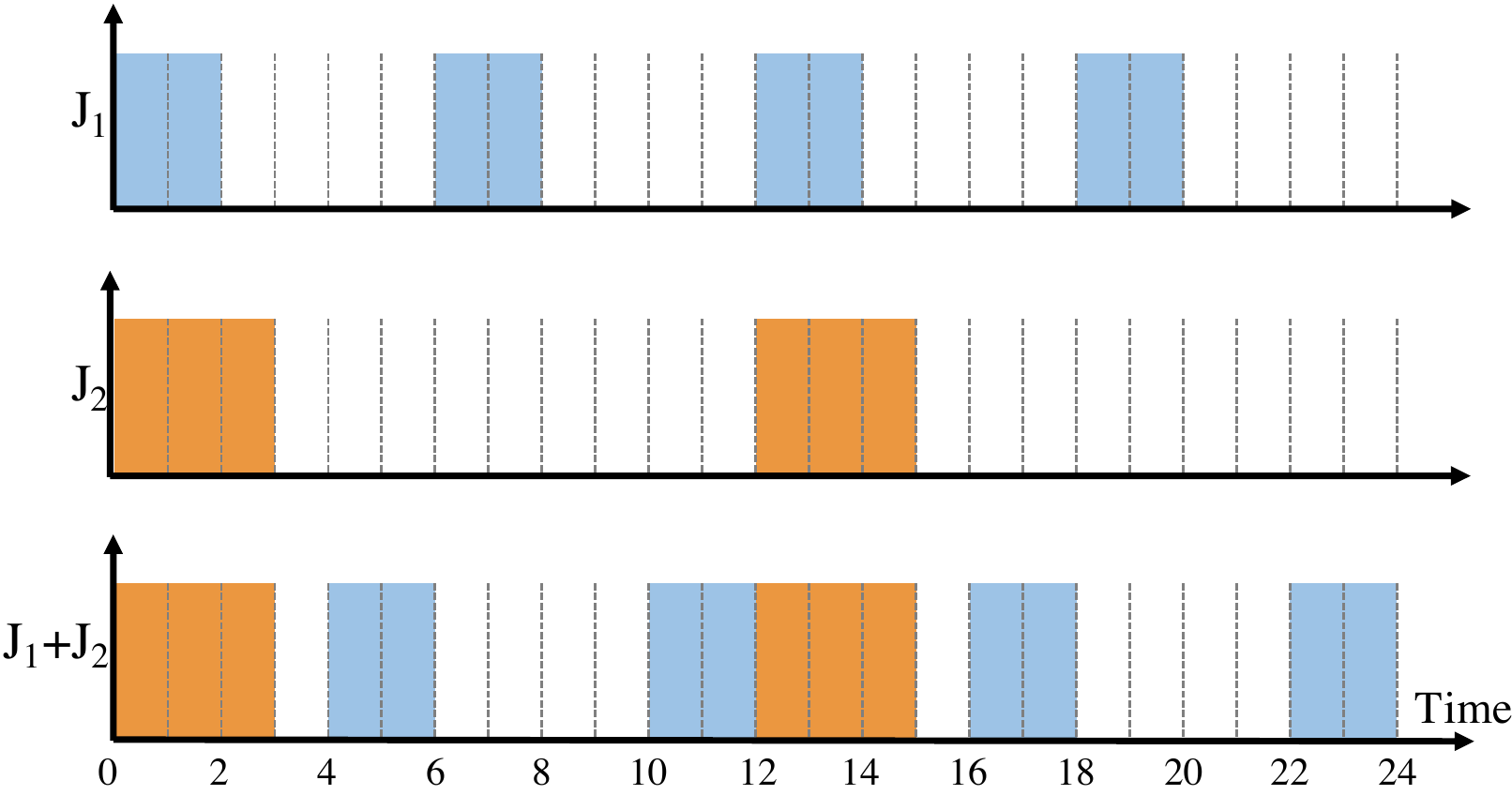}
  \caption[Toy examples of cyclic execution of {\pserver}.]{Toy examples of cyclic execution of {\pserver}.
  In the two figures at the top, the {\pserver} only serves one job ($J_1$ or $J_2$).
  In the bottom one, it serves both $J_1$ and $J_2$.
  The iteration duration of $J_1$ ($J_2$) is $6$ ($12$) units of time; 
  its model aggregation takes $2$ ($3$) units of time. 
  }
  \label{a_fig:toy-pack}
\end{figure}

After collecting the characteristics of a new training job, 
{\pservice} needs to assign the model aggregations from the temporary {\pserver}s to the stable ones and allocate new ones if needed. 
The objectives here includes minimizing the total number of {\pserver} for resource efficiency, 
and balancing the load of each {\pserver} for less resource contention among model aggregations

It is infeasible to tackle this assignment problem at scale in an online manner. 
When many DDL jobs are served by {\pservice},
the large number of tensors (parameters) in DL models and fast training speed
generate high volume of model aggregation requests\footnote{Each model aggregation task has one request per iteration.} per second,
which can easily overwhelm the scheduler and lead to non-negligible queuing delay.
Based on the periodicity of model aggregation workload, 
we propose an offline approach that gets 
fixed and cyclic execution slots for model aggregations.

\vspace{-12pt}
\paragraph{Cyclic Execution.} {\pserver} has an execution cycle 
that covers the execution slots of model aggregation tasks assigned to it.
The execution cycle is determined by the model aggregations on it and is updated when the assignments change.
When {\pserver} serves model aggregations from one job,
its execution cycle is simply the job's (also those model aggregations') iteration duration.
When model aggregations from multiple jobs are packed together,
{\pserver} picks the largest iteration duration among those jobs as its execution cycle.
With that, all the model aggregations can be executed within a single cycle.
And the jobs with smaller iteration duration may gets executed for multiple iterations.
In the toy examples of Figure~\ref{a_fig:toy-pack}, 
the execution cycle of {\pserver} is $6$ ($12$) units of time 
when only tasks of $J_1$ ($J_2$) are assigned to it.
If the tasks of $J_1$ and $J_2$ are packed together,
{\pserver} will have an execution cycle for $12$ units of time.
The model aggregations from $J_1$ will be executed twice within one cycle.

With this cyclic execution design, assigning a model aggregation task could 
change the execution cycle of {\pserver}, which affects the execution of existing tasks.
For example, there is an {\pserver} serving a model aggregation task 
whose execution time (iteration duration) is $1$ ($5$) unit of time.
If the task of $J_2$ in Figure~\ref{a_fig:toy-pack} is assigned to it, 
then its execution cycle will be $12$ units of time.
Accordingly, the iteration duration of the existing task will be $6$ units of time, 
since the task can run twice in one cycle.
Theoretically, the job of the existing task may 
lose $17$\% of its training speed.

\begin{table}[!t]
  \centering
  \small
  \caption{Notations in the assignment scheme.}
  \begin{tabular}{cl}
    \toprule
    Notation & Description \\
    \midrule
    $C_n$ & Execution cycle of {\pserver} $n$\\ 
    $C_n^{est}$ & Estimated $C_n$\\ 
    $D_j$ & Profiled iteration duration of job $j$ \\
    $d_j$ & Current iteration duration of job $j$\\
    $F_n^{est}$ & Estimated free CPU slots on Aggregation $n$\\
    $\mathbb{J}_n$ & Set of jobs that have tasks on {\pserver} $n$\\
    $\mathbb{T}_j^n$ & Set of tasks on {\pserver} $n$ that belong to job $j$\\
   \bottomrule
  \end{tabular}
  \label{a_tab:problem-notion}
\end{table}

With the purpose of eliminating potential performance loss, 
the assignment problem can be expressed as an integer programming problem (IP)
with the objective of minimizing the performance loss of all job (see Appendix).
Due to the non-linear constraints and objective function, the problem is {NP-hard} 
and is infeasible to solve.

Here, we introduce a heuristic-based solution (Pseudocode~\ref{alg:packing}).
The first step of our scheme
is to estimate the performance impact to the co-located jobs on each {\pserver} 
(Line~\ref{alg1:estimate_start} - \ref{alg1:estimate_end}).
Assuming the new task ($t$) is assigned to {\pserver} $n$,
it updates the execution cycle ($C_n^{est}$) of $n$ (Line~\ref{alg1:update_exe_cycle}),
and then, the iteration duration ($D_j^{est}$) of jobs on $n$ (Line~\ref{alg1:update_iter_duration}).
When any job's (estimated) performance loss exceeds the predefined threshold ($\texttt{LossLimit}$, default is $0.1$),
{\pserver} $n$ will be remove from the list of assignment destination (Line~\ref{alg1:remove_server}). 
After examining all the {\pserver}s, if no {\pserver} remains, the scheme will allocate a new one 
(Line~\ref{alg1:alloc_new_1}) and assign the task there (Line~\ref{alg1:assign_to_new_1}).
Meanwhile, how many free CPU slots ($F_n^{est}$) on each {\pserver} is calculated  
with the updated execution cycle and task execution (Line~\ref{alg1:free_cpu}).
Among the qualified {\pserver}s, our scheme assigns the new task to the best-fit one
who has sufficient but the least number of free CPU slots.
(Line~\ref{alg1:best_fit_start} - \ref{alg1:best_fit_end}).
In the end, if no one has enough free CPU to fit the new task,
a new {\pserver} will be allocated (Line~\ref{alg1:alloc_new_2} - \ref{alg1:assign_to_new_2}).
After assigning model aggregations to {\pserver}s, 
{\pservice} monitors the performance (\ie, training speed) of training jobs and compares with 
their standalone performance which is profiled at the beginning.
If there is any performance loss that exceeds the threshold (\texttt{LossLimit}), 
the new assignments will be reverted.

\floatname{algorithm}{Pseudocode}
\begin{algorithm}[!t]
  \small
  \begin{algorithmic}[1]
\INPUT{$t$ is new model aggregation task of job $k$} \newline
\hspace*{0.8em} $e_t$ is the execution (CPU) time of $t$ \newline
\hspace*{0.8em} $\mathbb{N}$ is the set of allocated {\pserver}s
  \FORALL{{\pserver} $n \in \mathbb{N}$} 
  \alglinelabel{alg1:estimate_start} 
    \STATE $C_n^{est} \gets$ max($C_n$, $D_k$) \alglinelabel{alg1:update_exe_cycle}
    \FORALL{job $j \in \mathbb{J}_n$}
      \STATE $d_j \gets$ max($D_j$, $\frac{C_n^{est}}{\floor*{\frac{C_n^{est}}{D_j}}}$) 
      \alglinelabel{alg1:update_iter_duration} %
    \ENDFOR
    \IF{$\frac{d_j - D_j}{d_j} \geq \texttt{LossLimit}$}
      \STATE $\mathbb{N} \gets \mathbb{N} \setminus n$, and skip 
      \alglinelabel{alg1:remove_server} %
    \ENDIF
    \STATE $F_n^{est} \gets C_n^{est} - \sum\limits_{j \in \mathbb{J}_n}(\floor*{\frac{C_n^{new}}{d_j}} \times \sum\limits_{i \in \mathbb{T}_j^n}e_i)$
    \alglinelabel{alg1:free_cpu}
  \ENDFOR
  \alglinelabel{alg1:estimate_end} 

  \IF{$\mathbb{N}$ is $\emptyset$}
    \STATE \textsl{Allocate} {\pserver} $s$
    \alglinelabel{alg1:alloc_new_1}
    \STATE $\mathbb{T}_k^s \gets t$, $\mathbb{N} \gets \mathbb{N} \cap s$
    \alglinelabel{alg1:assign_to_new_1}
    \STATE \textbf{return} $s$ and $\mathbb{N}$
  \ENDIF
  \vspace{4pt}

  \FOR{{\pserver} $n \in \mathbb{N}$ in descending order}
  \alglinelabel{alg1:best_fit_start}
    \IF{$F_n^{est} \geq e_t$ and is the best fit}
      \STATE $\mathbb{T}_k^n \gets \mathbb{T}_k^n \cap t$
      \STATE \textbf{return} $n$ and $\mathbb{N}$
    \ENDIF
  \ENDFOR
  \alglinelabel{alg1:best_fit_end}
  \vspace{4pt}

  \STATE \textsl{Allocate} {\pserver} $s$
  \alglinelabel{alg1:alloc_new_2}
  \STATE $\mathbb{T}_k^s \gets t$, $\mathbb{N} \gets \mathbb{N} \cap s$
  \alglinelabel{alg1:assign_to_new_2}
  \STATE \textbf{return} $s$ and $\mathbb{N}$
\end{algorithmic}
\caption{Model Aggregation Assignment Scheme}
\label{alg:packing}
\end{algorithm}

\vspace{-12pt}
\paragraph{Handling Outliers in Cyclic Execution.}
Due to random reasons (\eg, cache misses, and network variations),
workers in a DDL training job may become transient stragglers~\cite{nips13-ssp, aaai15-ssp, socc16-straggler},
which makes some model aggregation requests miss their execution slots in the execution cycle.
{\pservice} handles those delayed outliers in two different ways.
When a request arrives late, {\pserver} will check 
whether it has sufficient CPU slots for this delayed request 
after reserving enough slots for the remaining scheduled requests in current cycle.
If so, then the outlier will get executed.
Otherwise, the request will be postponed to the next cycle, 
so that the co-located model aggregations are not affected.
In worst case, the job could be delayed by one iteration.

\vspace{-6pt}
\subsubsection{{\pserver} Scaling}
\label{a_sec:elastic}
\vspace{-4pt}
Other than the model aggregation assignment scheme, 
{\pservice} needs to scale up or down the number of {\pserver}s 
to balance the tradeoff between CPU utilization and job performance.
{\pserver} scaling can be triggered by two events: job arrival and exit.

When a new job arrives, all of its model aggregations will be assigned 
by the scheme (\S\ref{a_sec:assignment}).
After that, if its performance is worse than the standalone one, 
{\pservice} will add a new {\pserver} and re-assign the entire job.
This procedure will repeat until the performance loss of job is within the threshold ($\texttt{LossLimit}$).

{\pservice} opportunistically recycles some light-loaded {\pserver}s 
when there are {\pserver}s released because of job exit.
Starting from the least-loaded {\pserver}, {\pservice} reassigns its workload 
to other {\pserver}s \emph{without} new allocations allowed.
If it succeeds, {\pservice} will recycle that {\pserver} and repeat the procedure 
on the next least-loaded one.

\subsubsection{{\pserver} Cluster}
\label{a_sec:scalability}
\vspace{-5pt}
It is a common issue that having a single centralized resource manager (\ie, {\psmaster}) may hurt the performance of system at scale.
In {\pservice}, assigning one model aggregation needs {\psmaster} to scan all available {\pserver}s in the pool.
One training job may have hundreds or even thousands of model aggregations (\ie, tensors) in its model.
Assuming {\pservice} gets deployed in a large-scale cluster, 
the time complexity of assigning one training job will exponentially increase with large numbers of {\pserver}s and model aggregations.
Additionally, it could take even longer if there are new {\pserver} allocations triggered in the assignments.
Therefore, {\psmaster} can be easily overwhelmed by a burst of job arrivals.
Moreover, on-demand resource scaling in {\pservice} could affect the execution of running jobs through workload reassignments.
A sequence of job events may thrash the workload assignments in {\pservice},
which makes it hard to get stable execution for the running jobs.

\begin{figure}[!t]
  \centering
  \includegraphics[scale=0.3]{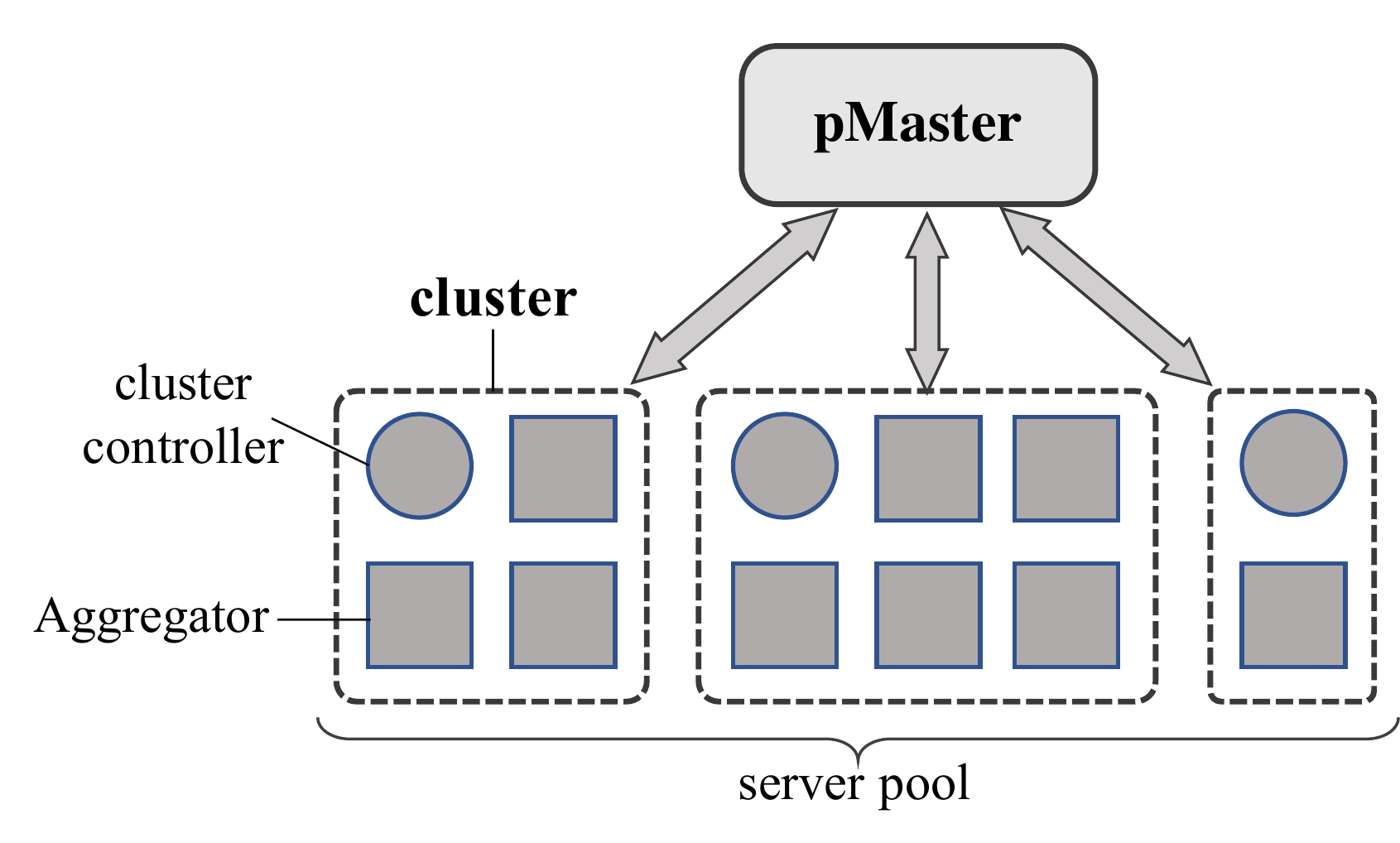}
  \caption[Overview of {\pserver} cluster.]{
    Overview of {\pserver} cluster.
  }
  \label{a_fig:cluster}
\end{figure}

To handle this issue in {\pservice}, we apply the idea of server (\ie, {\pserver}) cluster  
that the pool of {\pserver}s is split into multiple independent clusters (Figure~\ref{a_fig:cluster}).
Each cluster has a controller who is in charge of managing the {\pserver} resource in its cluster.
All cluster controllers are under the management of {\psmaster}.
Workload assignment is split into two steps.
A new job is firstly forwarded from {\psmaster} to a cluster controller, 
then the controller assigns model aggregations of the job to its {\pserver}s.
Therefore, one job gets {\pserver} resource from a single cluster.
Each cluster works independently.

\vspace{-12pt}
\paragraph{Why {\pserver} Cluster?}
Splitting the pool of {\pserver}s into multiple clusters brings the following benefits in mitigating the scalability issue.
First, it is the cluster controller that allocates {\pserver} resource to model aggregations.
The number of assignment destinations in a cluster is much fewer than the total number of {\pserver}s in {\pservice}.
Second, model aggregations of a job get the {\pserver} resource from a single cluster.
Only the jobs that are served by the same {\pserver} cluster may be affected by a job event (arrival or exit).
The impact of workload reassignments is limited within a single cluster.
Moreover, multiple cluster controllers could perform workload assignments in parallel when there are multiple job arrivals.

With {\pserver} cluster design, {\pservice} assigns model aggregations of new jobs in two steps.
First, {\psmaster} decides which cluster should be chosen for placing the new job.
{\psmaster} keeps track of the remaining free CPU resource of each {\pserver} cluster.
According to the profiled job information, specially total CPU consumption of the job, 
{\psmaster} selects the best-fit cluster, who has sufficient but least amount of free CPU resource, and forwards the new job there.
The complexity of this step is negligible compared to the assignment scheme in Pseudocode~\ref{alg:packing}.
Once the job is forwarded to a cluster, the cluster controller assigns model aggregations in this job to its {\pserver}s 
following the assignment scheme in \S\ref{a_sec:assignment}.
As discussed above, the time complexity in this step is greatly reduced because of much fewer assignment destinations.
When a job exits, its cluster controller should report this event to {\psmaster}.
{\psmaster} will update the status of the cluster on its side for future job forwarding.

\vspace{-12pt}
\paragraph{Hybrid Resource Scaling.}
As mentioned above, on-demand resource scaling triggered by job arrivals or exits may jeopardize the execution of running jobs 
when there are too many workload reassignments generated.
Although periodically resource scaling could reduce such disruptions, 
it can not respond to the change of resource demand in real-time. 
Consequently, {\pservice} performs resource scaling in a hybrid manner.
There is a predefined resource scaling period in {\pservice}.
{\pservice} adjusts the number of clusters and the amount of {\pserver}s in each cluster
according to the resource demand measured in the latest period.
To avoid starvation, on-demand {\pserver} allocation is also allowed 
when the demand of new {\pserver} is higher than a threshold.

There are interactions between {\psmaster} and cluster controllers when job events or resource scaling occurs.
For a new job arrival, {\psmaster} needs to forward the new job information to the chosen cluster controller for assigning model aggregations.
When a job exit, the cluster controller sends job completion information to {\psmaster} for bookkeeping.
A cluster controller should send allocation or deallocation requests to {\psmaster} when resource scaling is triggered by job events in it.
Approval has to be received before the cluster controller can carry out the operation.
The amount of those interaction messages is at the same order of magnitude of job arrival and exit rate.
Therefore, these interactions will not be the bottleneck of {\pservice}.
\vspace{-4pt}

\section{Implementation}
\vspace{-6pt}
We have implemented the design of {\pservice} into a system named {\autops}.
{\autops} is built on top of {ps-lite}~\cite{osdi14ps} with about $5$K lines of C++ code added.

In {\autops}, {\psmaster} is a daemon process that serves the entire cluster.
It opens a connection address to the DDL training jobs who want to use {\pservice}.
A training job can connect to {\psmaster} through the {\psagent}s that are collocated with its workers.
{\psagent} is a implemented as a Key-Value library that is loaded by the DL framework of worker.
{\psagent} exposes the standard \emph{Push} and \emph{Pull} APIs to the Key-Value store layer in DL framework for model aggregation requests and responses.
Since RDMA network is widely deployed in many DL clusters, our current implementation uses RDMA \texttt{ibverbs} for communication.
Every {\psagent} and {\pserver} has a control channel to the {\psmaster}.
{\psagent} needs to send requests to {\psmaster} for job registration and tensor initialization.
In response, {\psmaster} sends back the initial assignments of tensor.
When re-assigning workload, {\psmaster} sends the migration command to the {\pserver} that is currently hosting the tensor, and waits completion notifications from the two related {\pserver}s.
In addition, {\psmaster} sends profiling commands to {\psagent}s and {\pserver}s when there are updates in workload assignment or {\pserver} allocation.

\section{Evaluation}
\label{a_sec:eval}
\vspace{-5pt}

We evaluated {\autops} in testbed experiments and trace-driven simulations, and highlight the results as follows:
\begin{denseitemize}
  \item {\autops} improves the resource (CPU) efficiency with none or negligible performance loss to the training jobs.
  It can reduce up to $75\%$ of CPU servers comparing to the traditional parameter server approach.

  \item {\autops} outperforms the traditional parameter server approach (up to $1.17\times$) when a single job uses each of them in standalone mode.

  \item The CPU saving benefits of {\autops} hold for the large-scale cluster scenario in trace-driven simulation.

\end{denseitemize}

\begin{figure}[!t]
  \centering
  \includegraphics[scale=0.35]{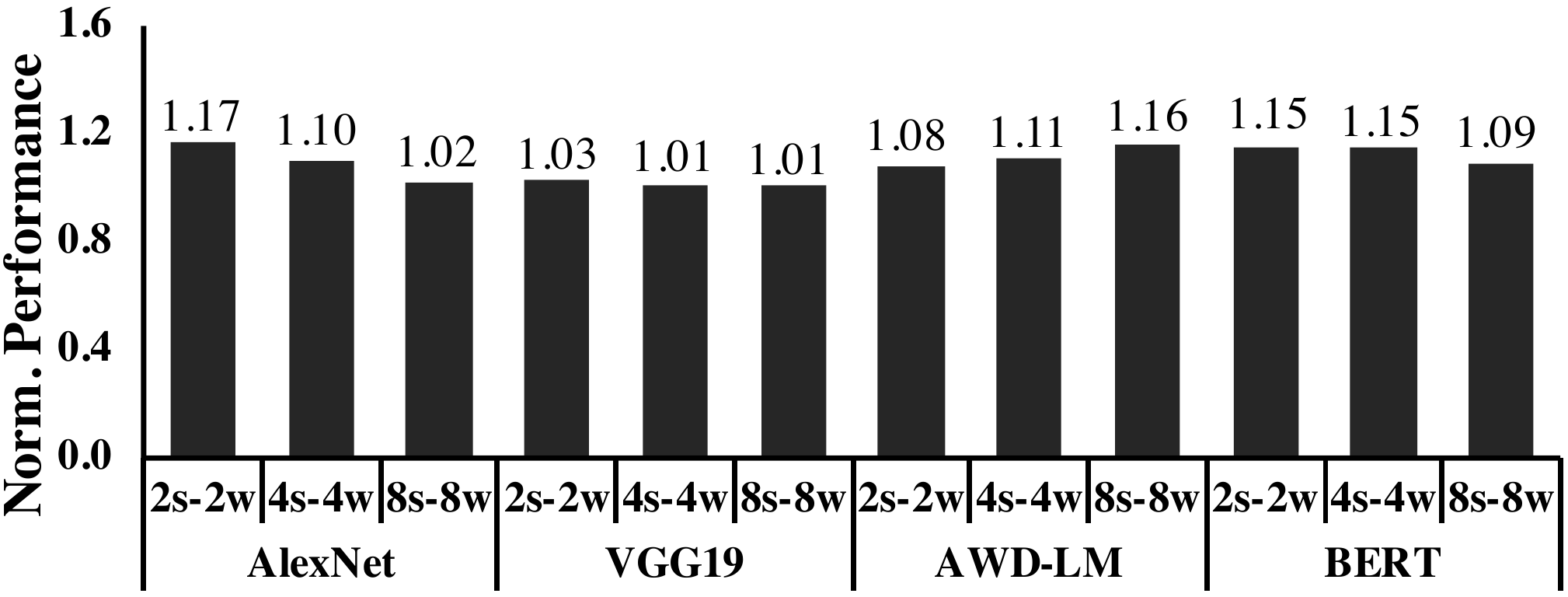}
  \caption[Normalized performance of single job using {\autops}.]{Normalized performance of single job using {\autops}.
  The performance of each job is normalized by its performance when using {ps-lite}.
  For each job, it use the same number of {\pserver}s ({\autops}) and parameter servers ({ps-lite}).
  }
  \label{a_eval:single-perf}
\end{figure}

\vspace{-4pt}
\subsection{Experimental Setup}
\vspace{-4pt}
\paragraph{Testbed.}
Our testbed consists of $8$ GPU machines and $8$ CPU machines. 
All machines connect to a $100$ Gbps RDMA network.
Each GPU machine has $4$ {NVIDIA} {Telsa} {P100} GPUs with {NVLink} connections.

\vspace{-8pt}
\paragraph{Workload.}
We train $4$ classic and popular DL models%
using {\mxnet} in the experiments.
Two of them are CNN models ({\alexnet}~\cite{alexnet-nips12}, {\vgg19}~\cite{vgg-14}), 
which are trained with {\imagenet}~\cite{imagenet-cvpr} dataset.
The other two are RNN models ({\lm}~\cite{lm-17}, {\bert}~\cite{bert-google}).
{\lm} is trained with {\wikidata}~\cite{wikitext-2} dataset; {\bert} uses {\bertdata}~\cite{bookscorpus-iccv15}.
Batch size of each model is set to the maximum to fit into the GPU memory in our testbed.
In this chapter, ``1s-2w" means the job requires $1$ parameter server and $2$ workers.
Parameter server (and {\pserver}) uses CPU machine, and worker runs on GPU machine.
Each worker has $4$ GPUs (from the one machine).

\vspace{-6pt}
\paragraph{Baseline.}
We compare {\autops} to the classic parameter server implementation ({ps-lite}~\cite{osdi14ps}, also using RDMA network).
Using {ps-lite}, each training job has an individual group of parameter servers for model aggregation.
{\autops} also takes the number of parameter servers of each job to profile the standalone performance of the job.

\vspace{-6pt}
\paragraph{Simulator.}
We build an event-based simulator and use a real job trace from {Microsoft} to evaluate {\autops} when it gets deployed on a large scale cluster.
It simulates all job events and resource scaling activities.

\vspace{-6pt}
\paragraph{Metric.}
We use the training speed (\ie, samples per second) to represent job performance.
{\autops} saves CPU resource through reducing the number of {\pserver}s allocated for model aggregation.
In testbed experiments, we focus on how many CPU servers are reduced in {\autops} comparing to {ps-lite} under the same scenarios, and define:
\begin{equation*}
  \text{CPU Reduction Ratio} = \frac{\text{\# of param. servers $-$ \# of Agg.}}{\text{\# of param. servers}}
\end{equation*}

\vspace{-5pt}
Larger value of this ratio means savings more CPU servers for model aggregation.

\vspace{-5pt}
\subsection{Evaluation Results}
\label{a_sec:experiment}
\vspace{-4pt}

\subsubsection{Single-Job Experiments}
\label{a_sec:single-job-exp}
\vspace{-4pt}
To verify the effectiveness of model aggregation function in {\autops},
we compare the performance of jobs when they are using {\autops} and {ps-lite} in standalone mode (Figure~\ref{a_eval:single-perf}).

The performance of {\autops} is not worse than {ps-lite}, which means the extra operations 
(\eg, extra request mappings for decoupling model aggregation, and periodic job profiling) have negligible impact on the training job.
In addition, {\autops} outperforms {ps-lite} by up to $1.17\times$ in some cases.
These performance improvements come from the better balanced load distribution in {\autops},
comparing to the round-robin distribution in {ps-lite}.

\vspace{-4pt}
\subsubsection{Multi-Job Experiments}
\label{a_sec:multi-job-exp}
\vspace{-4pt}
When multiple jobs run on {\autops}, it can opportunistically shrink the number of allocated {\pserver}s for resource efficiency.
Here, we run multiple training jobs (with the same DL model and distributed settings) together to see how jobs' performance and CPU server allocation will be changed.
Due to the limitation of machines, we can run up to $4$ (2s-2w) jobs, or $2$ (4s-4w) jobs.  

\begin{figure}[!t]
  \centering
  \subfloat[][\alexnet]{
      \includegraphics[scale=0.3]{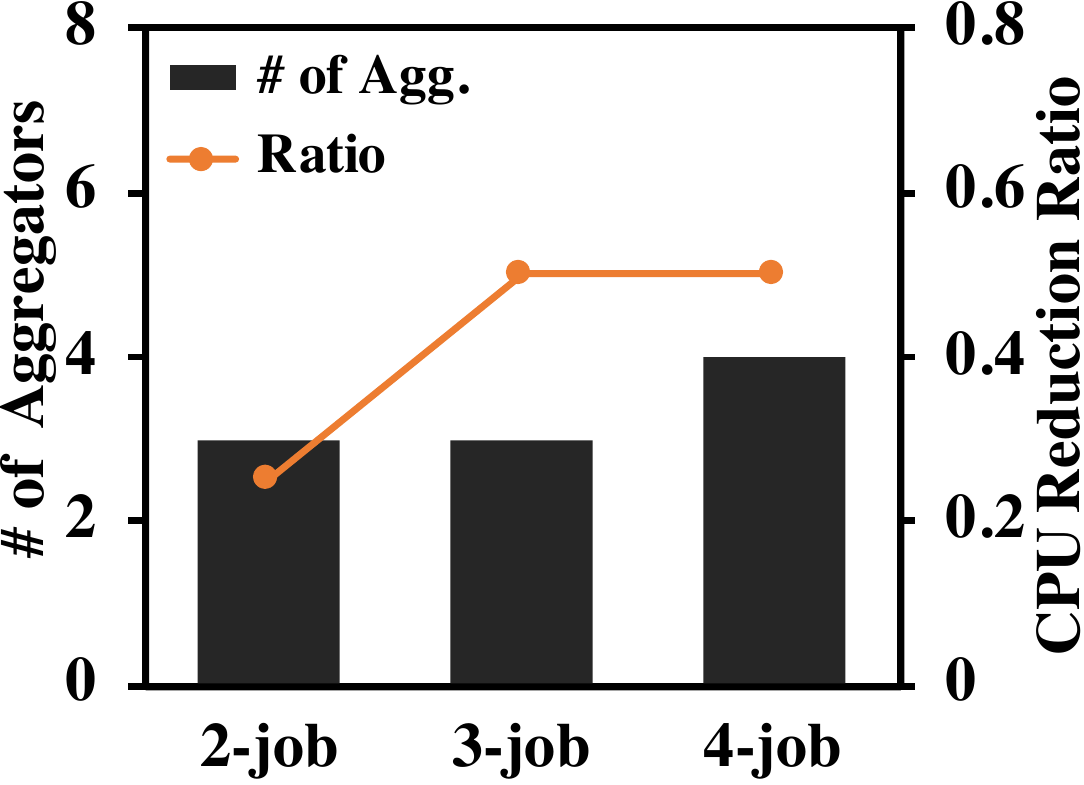}
      \label{a_eval:alex-server}
  }
  \subfloat[][\vgg19]{
      \includegraphics[scale=0.3]{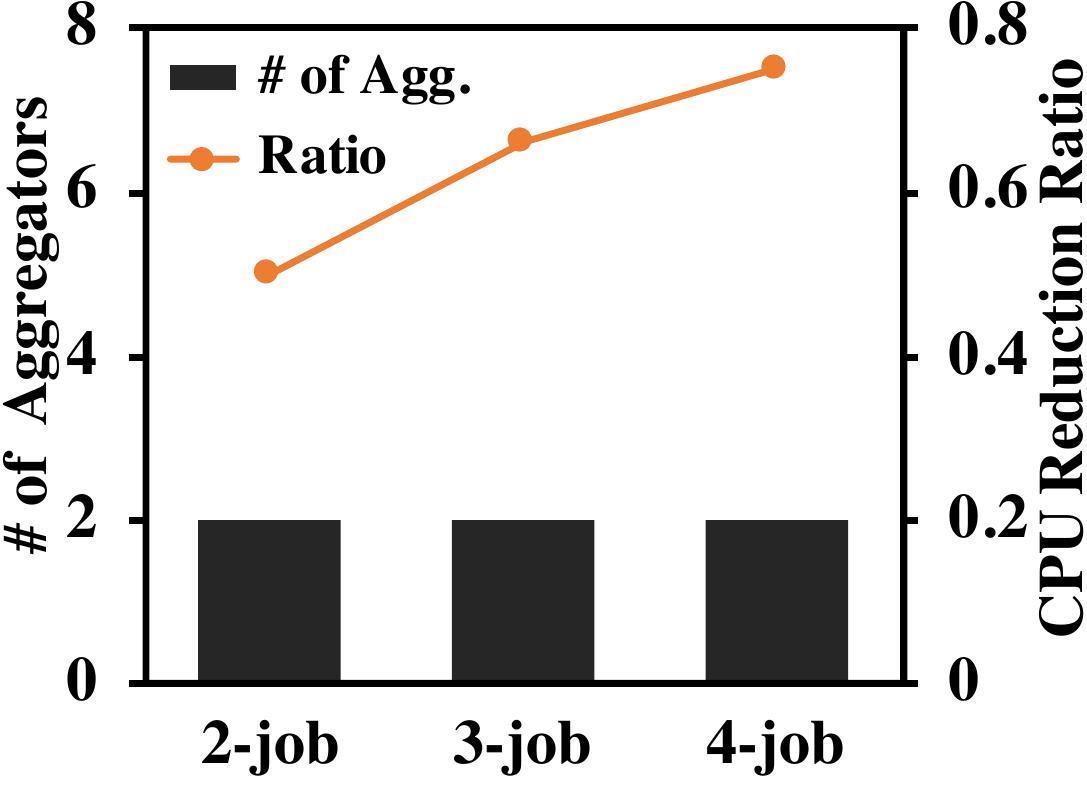}
      \label{a_eval:vgg-server}
  }
  \\
  \subfloat[][\lm]{
      \includegraphics[scale=0.3]{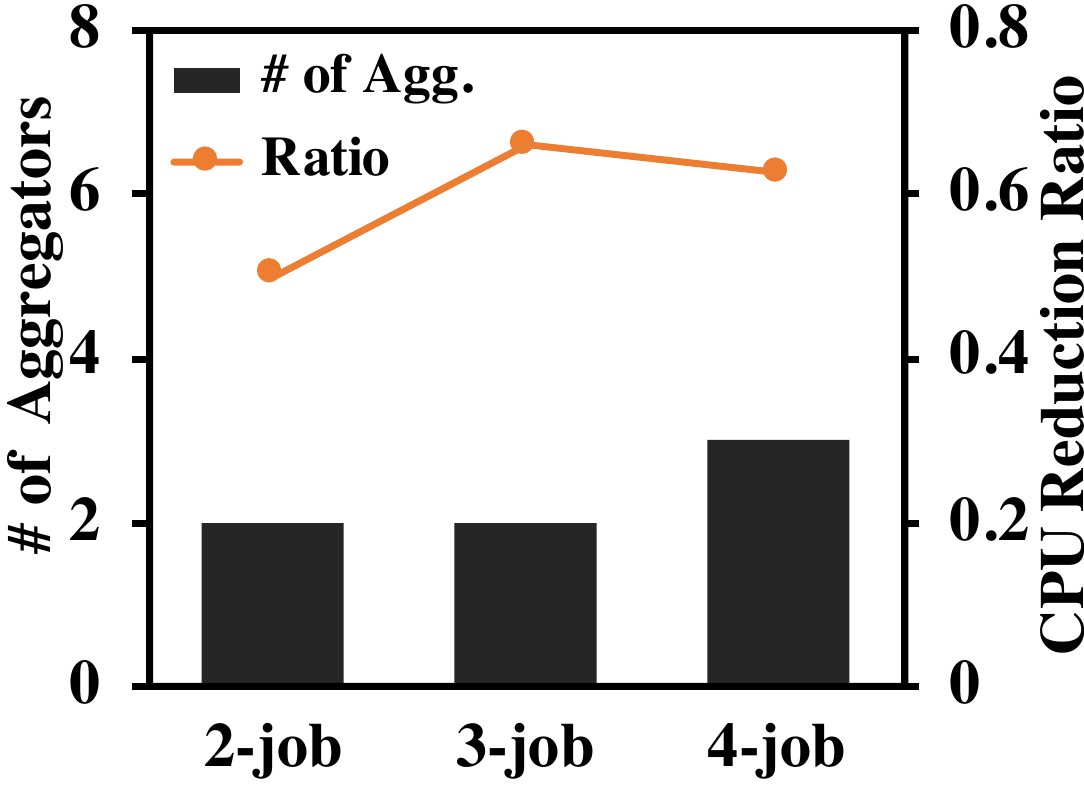}
      \label{a_eval:lm-server}
  }
  \subfloat[][\bert]{
      \includegraphics[scale=0.3]{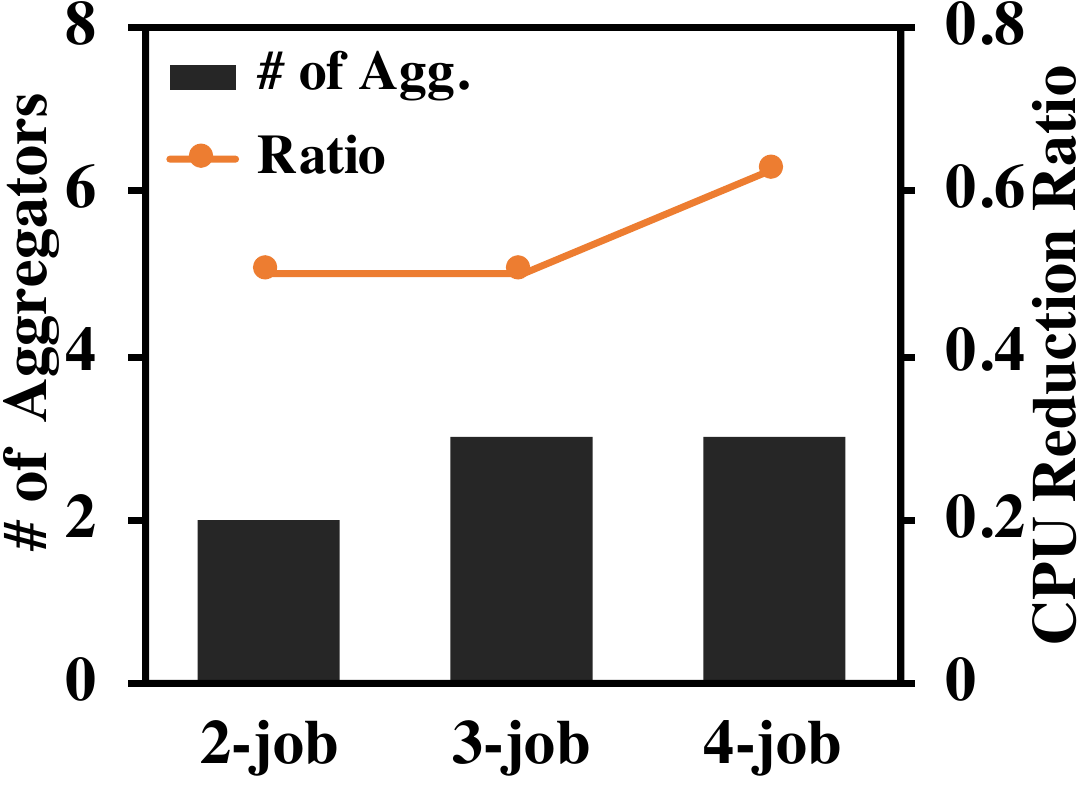}
      \label{a_eval:bert-server}
  }
  \caption[Number of {\pserver}s when multiple (2s-2w) jobs use {\autops} together.]{Number of {\pserver}s when multiple (2s-2w) jobs use {\autops} together.
  Each job requires $2$ parameter servers when using {ps-lite}.
  $2$-job means two jobs use {\autops} together.
  Jobs in the same group train the same model.
  }
  \label{a_eval:2s-2w-server}
\end{figure}

\begin{table}[!t]
  \centering
  \small
  \caption{CPU reduction ratio when $2$ (4s-4w) jobs use {\autops}.}
  \begin{tabular}{rcccc}
    \toprule
    & \alexnet & \vgg19 & \lm & \bert \\
    \midrule
    Ratio & $0.375$ & $0.5$ & $0.5$ & $0.5$ \\ 
  \bottomrule
  \end{tabular}
  \label{a_eval:4s-4w-server}
\end{table}

\vspace{-10pt}
\paragraph{CPU Reduction.}
Figure~\ref{a_eval:2s-2w-server} shows the number of allocated {\pserver}s in {\autops} when multiple (2s-2w) jobs runs together.
The baseline here is the total number of required parameter servers in each scenario.
For example, there are $6$ parameter servers needed in total when $3$ (2s-2w) jobs use {ps-lite} for model aggregation.
Comparing to {ps-lite}, {\autops} can save $25\%$ to even $75\%$ of CPU servers.
The {\alexnet} jobs need more {\pserver}s than the jobs of other models.
{\alexnet} is the only model that requires one extra {\pserver} to run $2$ (2s-2w) jobs.
That's because of the very short iteration time of {\alexnet} jobs,
which makes them have much higher frequency of model aggregation than others.
In contrast, $2$ {\pserver}s can serve $4$ {\vgg19} jobs whose iteration time is much longer.
Table~\ref{a_eval:4s-4w-server} shows the CPU reduction ratio of {\autops} when there are $2$ (4s-4w) jobs.
Same as Figure~\ref{a_eval:2s-2w-server}, most of the ``$2$-job" cases can run without allocating new {\pserver}.

\begin{figure}[!t]
  \centering
  \includegraphics[scale=0.35]{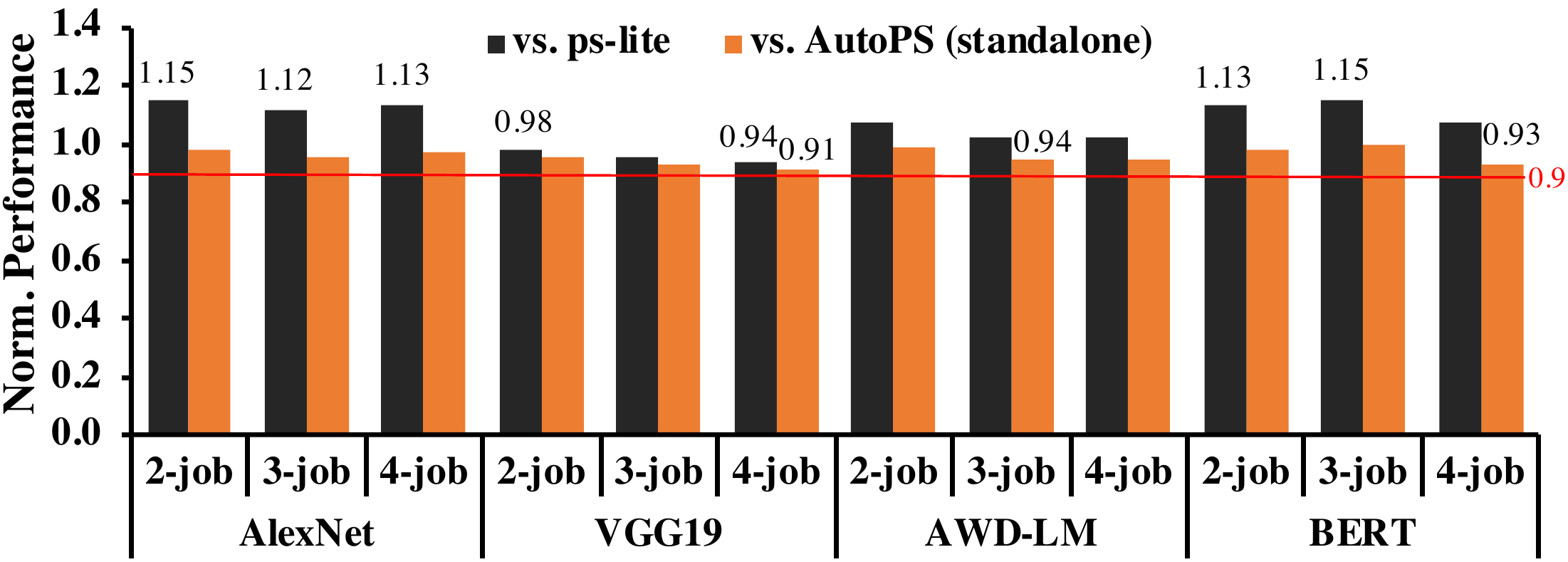}
  \caption[Performance impact when multiple (2s-2w) jobs share {\autops}.]{Performance impact when multiple (2s-2w) jobs share {\autops}.
  Jobs in the same group train the same model.
  The averaged performance of each multi-job group is used.
  For comparison, it is normalized by the performance of the same job when using {ps-lite} and {\autops} in standalone mode, respectively.
  }
  \label{a_eval:multi-job-perf}
\end{figure}

\vspace{-10pt}
\paragraph{Impact on Job Performance.}
In {\pservice}, the performance of training jobs should not be sacrificed for improving resource efficiency. 
When any workload assignment makes the job performance lower than the threshold (\lowperf), {\autops} will revoke it and re-do the assignment with new {\pserver} added.
Figure~\ref{a_eval:multi-job-perf} shows how job performance is impacted when multiple jobs uses {\autops}.
The number of allocated {\pserver}s of each multi-job group can be found in Figure~\ref{a_eval:2s-2w-server}.
The average performance is measured when all jobs in the group are in stable state.
Because of the performance protection (\lowperf), the performance loss caused by resource contentions among jobs are limited and even negligible in {\autops}.
Comparing to the performance from {\autops} (in standalone mode), sharing {\autops} among multiple jobs jobs may lose up to $9\%$ training speed in our experiments.
In some cases, the averaged performance of multiple jobs using {\autops} is even better than the standalone performance from {ps-lite}.

To conclude, {\autops} can reduce the number of {\pserver}s allocated for model aggregation through packing workload from multiple jobs. 
Meanwhile, it imports negligible performance loss to the training jobs.

\begin{figure}[!t]
  \centering
  \includegraphics[scale=0.30]{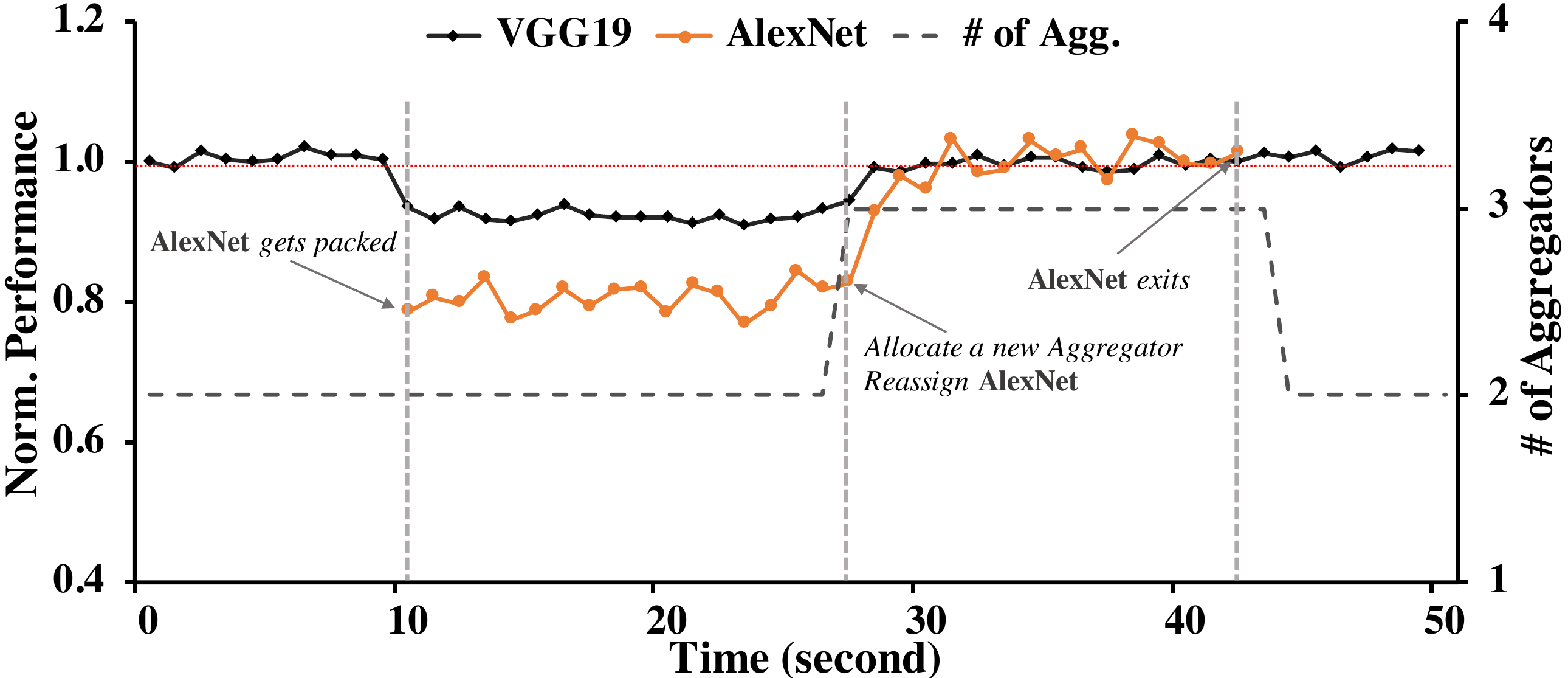}
  \caption[Time trace of the two-job case study.]{Time trace of the two-job case study.
  The training performance of the jobs is normalized by their standalone performance, respectively.
  }
  \label{a_eval:pack-time-trace}
\end{figure}

\vspace{-10pt}
\paragraph{Case Study of {\pserver} Scaling.}
We bring a case study with two jobs to show how {\autops} scales {\pserver} when job events (\ie, arrival and exit) occur.
Figure~\ref{a_eval:pack-time-trace} shows how the performance of jobs and {\pserver} allocation is changed when job events occur.
There is a {\vgg19} (2s-2w) job that uses {\autops} for model aggregation and is already in steady state.
Following its parameter server requirement, {\autops} allocates $2$ {\pserver}s for it.
A new {\alexnet} (2s-2w) job just completed its initial performance profiling, and gets its first assignments to the two existing {\pserver}s at $11^{th}$ second.
The performance of {\vgg19} job is slightly affected because of resource contentions on those two servers.
However, the new {\alexnet} job loses up to $22\%$ of its performance.
After monitoring enough iterations (default is 100), {\autops} determines the assignments of {\alexnet} job should be revoked.
At $27^{th}$ second, {\autops} allocates a new {\pserver} and reassigns the {\alexnet} job.
Both of the two jobs get better performance after that.
The {\alexnet} job completes at the $42^{nd}$ second. 
Since the newest {\pserver} only has the model aggregations from {\alexnet}, {\autops} releases it right after job exit.

\vspace{-5pt}
\subsubsection{Trace-Driven Simulation}
\vspace{-5pt}
We evaluate {\autops}'s performance in CPU saving using a real job trace.
This is 10-week job trace from a $2000$-GPU cluster in {Microsoft}~\cite{atc19-shared-cluster}.
We compare the CPU consumption of {\autops} against the CPU requirements of running jobs specified by users.
Because {ps-lite} allocates the required amount of CPU resource for each job if it is used for model aggregation.

\begin{figure}[!t]
  \centering
  \includegraphics[scale=0.28]{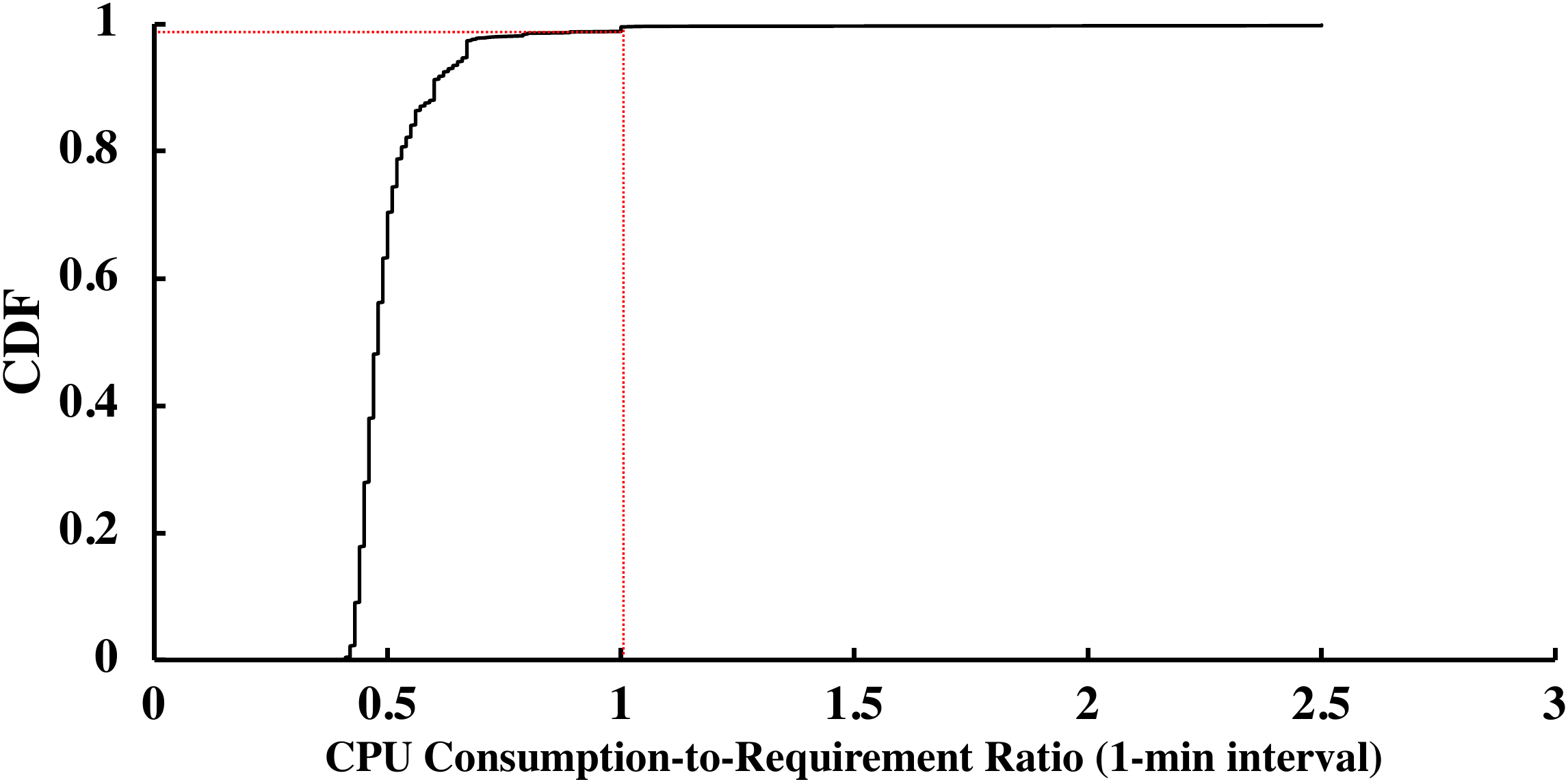}
  \caption[CPU consumption of {\autops} compared to total CPU requirements of running jobs in the trace-driven simulation.]{
    CPU consumption of {\autops} compared to total CPU requirements of running jobs in the trace-driven simulation.
    The x-axis is the ratio of allocated CPU cores of {\autops} to total CPU requirements of running jobs.
    CPU consumption and CPU requirements are measured with 1-min interval.
  }
  \label{a_eval:sim-cpu-saving}
\end{figure}

\vspace{-8pt}
\paragraph{CPU Savings.} 
We verify that the CPU saving benefits of {\autops} still hold when it gets deployed in a large-scale cluster.
Figure~\ref{a_eval:sim-cpu-saving} compares the CPU consumption of {\autops} against total CPU requirements of running jobs in the trace-driven simulation.
The x-axis in the figure is the ratio of allocated CPU servers of {\autops} to total CPU requirements of running jobs.
Smaller value of this ratio means more CPU savings of {\autops}.
Over $99$\% of the time, this ratio is lower than $1$, 
which means {\autops} could save CPU resource for model aggregation for majority of the time.
Very rarely, {\autops} consumes more CPU resource than the CPU requirements from users.
In worst cases, the ratio can be even larger than $2.5$. 
These come from the scenarios that some allocated CPU resource in {\autops} are idle because of recently-completed jobs.
Due the periodical resource scaling in \S\ref{a_sec:scalability},
{\autops} could not release the free CPU resource until the end of current period,
which makes CPU consumption of {\autops} being higher than the CPU requirements of running jobs.
Overall, {\autops} could reduce the CPU cost (\ie, CPU time) of model aggregations by $52.7$\% in the simulated scenario.

\vspace{-4pt}
\section{Discussion}
\label{a_sec:discussion}
\vspace{-3pt}
\paragraph{Utilizing Transient Resource.}
The cloud clusters usually have some amount of transient resource that has short available time. 
Cloud providers often sell the transient resource to users as spot instances with discounted price~\cite{spot-instance}. 
With resource elasticity in {\pservice}, it is feasible to run model aggregations on spot instances for cost saving purpose.
When the spot instance of a {\pserver} is going to expire, other {\pserver}s could immediately take over the affect model aggregations 
through workload reassignment with negligible overhead.

\vspace{-10pt}
\paragraph{Performance Isolation for Multitenancy.}
To pursue CPU efficiency, {\pserver} may assign model aggregations from different jobs to the same {\pserver} 
where jobs interfere with each other.
Using a feedback-based assignment scheme, {\pservice} limits performance degradation within a certain threshold ($\texttt{LossLimit}$).
However, this mechanism is infeasible to the multi-tenant environment where jobs may have different performance requirements.
{\pservice} needs a isolation scheme which can satisfy the requirements from different users.
\vspace{-5pt}

\vspace{-5pt}
\section{Related Work}

\vspace{-5pt}
\paragraph{Priority-based Model Aggregation.}
{ByteScheduler}~\cite{bytesheduler} is state-of-the-art data parallel training framework based on parameter server architecture.
It optimizes the execution order of model aggregations according to the execution DAG of the training job and 
enforces their priorities in the data communication layer for less queuing delay.
{\pservice} differs from {ByteScheduler} in two fundamental ways.
First, the role of parameter server in {ByteScheduler} has been changed.
The function of updating model parameters, which is originally executed by parameter server, is moved to the execution engine at worker side.
Therefore, the parameter server in {ByteScheduler} is just a hub that redistributes local gradients from workers.
Second, the two PS-based systems working on different problems.
{\pservice} focuses on reducing CPU consumption without sacrificing training performance.
{ByteScheduler}~\cite{bytesheduler} aims at improving training performance without considering CPU consumption.
Moreover, comparing to {ByteScheduler}, {\pservice} has two benefits that are originated from the design of decoupling model aggregation.
First, {\pservice} is easy to use since it is transparent to DL frameworks. 
Second, {\pservice} can utilize the transient resource in the cloud through its elastic feature.
{ByteScheduler} is incapable of dynamically changing its underlying resource.
Actually, the priority-based aggregation scheme in {ByteScheduler} can be integrated into {\pservice} to improve training performance.

\vspace{-10pt}
\paragraph{Alternatives for Model Aggregation.}
{Horovod}~\cite{horovod} applies bandwidth optimal ring-based {AllReduce} algorithms for 
model aggregation in DDL training.
This approach can fully utilize the network bandwidth among the training nodes.
But it requires homogeneous hardware (specially network links of equal bandwidth) 
which does not hold in the shared cluster.
{ParameterHub}~\cite{parameterhub} is a physical machine designed for model aggregation.
Similar to {\pservice}, {ParameterHub} provides a cluster-wise parameter hosting and aggregation function to multiple DDL training jobs.
However, it is not cost-effective because it can not flexibly scale-up or down according to change of its load.

\vspace{-9pt}
\paragraph{In-Network Model Aggregation}
With the trend of deploying programmable network devices in the cluster, in-network model aggregation has been proposed in the recent years.
{SwitchML}~\cite{switchml-nsdi21} uses programmable switch dataplane to execute model aggregations.
It reduces the volume of exchanged data during model aggregation and network latency for accelerating the training speed.
{iSwitch}~\cite{iswitch-isca19} is an in-switch aggregation acceleration solution for distributed reinforcement learning training.
Focusing on the smaller but more frequent aggregations in reinforcement learning training, 
it conduces network packet level aggregation rather than the entire gradient vectors to reduce the aggregation overhead.
{SHARP}~\cite{sharp-2016} is a collective technology from {Mellanox} that is commonly applied for in-network aggregation.
It's only available in certain {InfiniBand} switches and comes with fixed functions, which make it difficult to evolve to support new aggregation approaches.

\vspace{-9pt}
\paragraph{Elastic Scaling on GPU Workers.}
Applying resource elasticity in GPU workers %
could significantly improve GPU efficiency and shorter job completion time.
However, existing DL frameworks either apply a fixed number of GPU workers throughout the lifetime of jobs, 
or adjusts the number of workers with high overheads that counteracts the benefits from elasticity.
Recent work~\cite{andrew-mlsys2020, edl-2019} has been proposed to cut down the overhead of scaling GPU workers.
Furthermore, \cite{andrew-mlsys2020} has an autoscaling engine, which considers account cost, 
other than GPU efficiency and job performance.%
\cite{edl-2019} develops an elasticity-aware DL scheduler that achieves various scheduling objectives 
in multi-tenant GPU clusters.

\vspace{-10pt}
\section{Conclusion}
\vspace{-6pt}
{\pservice} is an elastic and shared model aggregation service for emerging DDL training jobs.
It could enhance the utilization of CPU resource for model aggregations without hurting job performance.
In {\pservice}, multiple jobs can submit their model aggregations to the unique interface without allocating and managing their own parameter servers. 
Internally, {\pservice} balances the tradeoff between resource efficiency and job performance through 
dynamic workload and elastic resource scaling.
Our implementation of {\pservice}, called {\autops}, saves up to $75$\% of CPU servers when serving DDL training jobs.
Its CPU saving benefits hold when it gets deployed in a large-scale cluster.
More importantly, {\pservice} is totally transparent to user and can easily be adopted by popular DL frameworks.
\label{lastpage}

{
\bibliographystyle{abbrv}
\bibliography{AutoPS}
}
\appendix
\section{Decoupled Model Aggregation}
\label{appx:decouple}
In order to manage model aggregations from multiple jobs, {\pservice} firstly decouples them from training jobs.

Originally, each training job maintains a group of parameter servers (PS) of its own.
The assignment of model aggregation is statically decided by local worker before the job starts.
Model aggregation requests use a \emph{Key-Value} format. 
The key field is filled with the tensor ID, so that the request can be easily identified at both worker and PS side.

\begin{figure}
    \centering
    \includegraphics[scale=0.45]{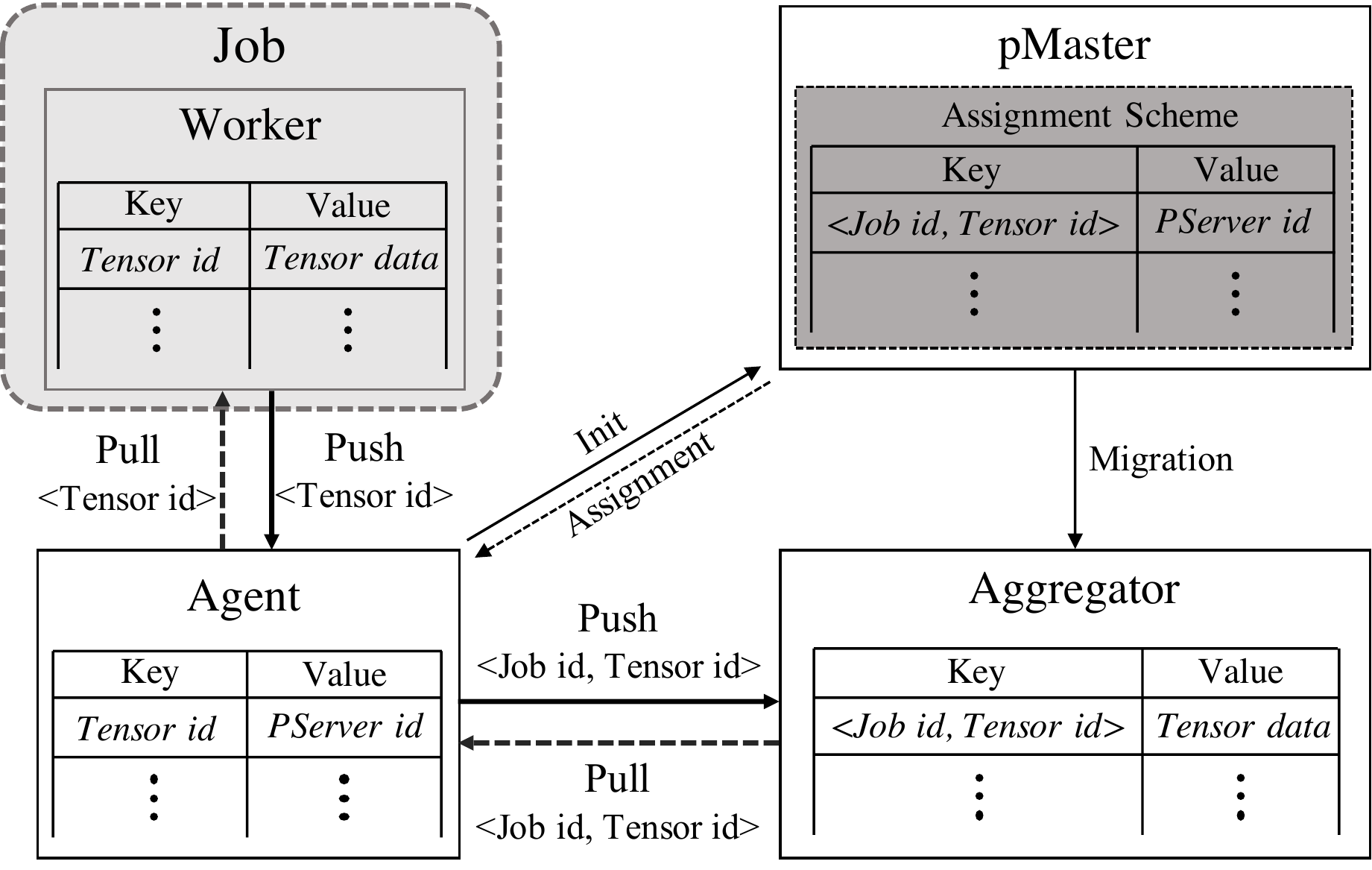}
    \caption{Interactions among worker and {\pservice} components.
    Model aggregation requests in {\pservice} are identified by the combination of job ID and tensor ID. 
    {\psagent} sends \emph{Init} to {\pservice} when a new tensor arrives.
    {\pserver} receives \emph{Migration} from {\pservice} when one of its tensor is reassigned.
    }
    \label{fig:interaction}
\end{figure}

As a major mechanism in {\pservice}, the function of workload assignment is moved from individual training job to {\psmaster} (Figure~\ref{fig:interaction}).
For each worker, there is one {\psagent} that has a mapping table for tensor assignment. It can assist the worker to figure out where the requests should be forwarded.
When an aggregation request of a new tensor arrives at {\psagent}, it will send the initial request to {\psmaster} for assignment.
Other than tensor ID, the job information (\ie, job ID) also has to be kept for each model aggregation request in {\pservice}, since an {\pserver} might be shared by multiple training jobs.
Therefore, the key field of aggregation request is transferred to be a pair of job ID and tensor ID by {\psagent} before sending to the destination.

\section{Tensor Migration}
\label{appx:migration}

\begin{figure}[!t]
    \centering
    \includegraphics[scale=0.54]{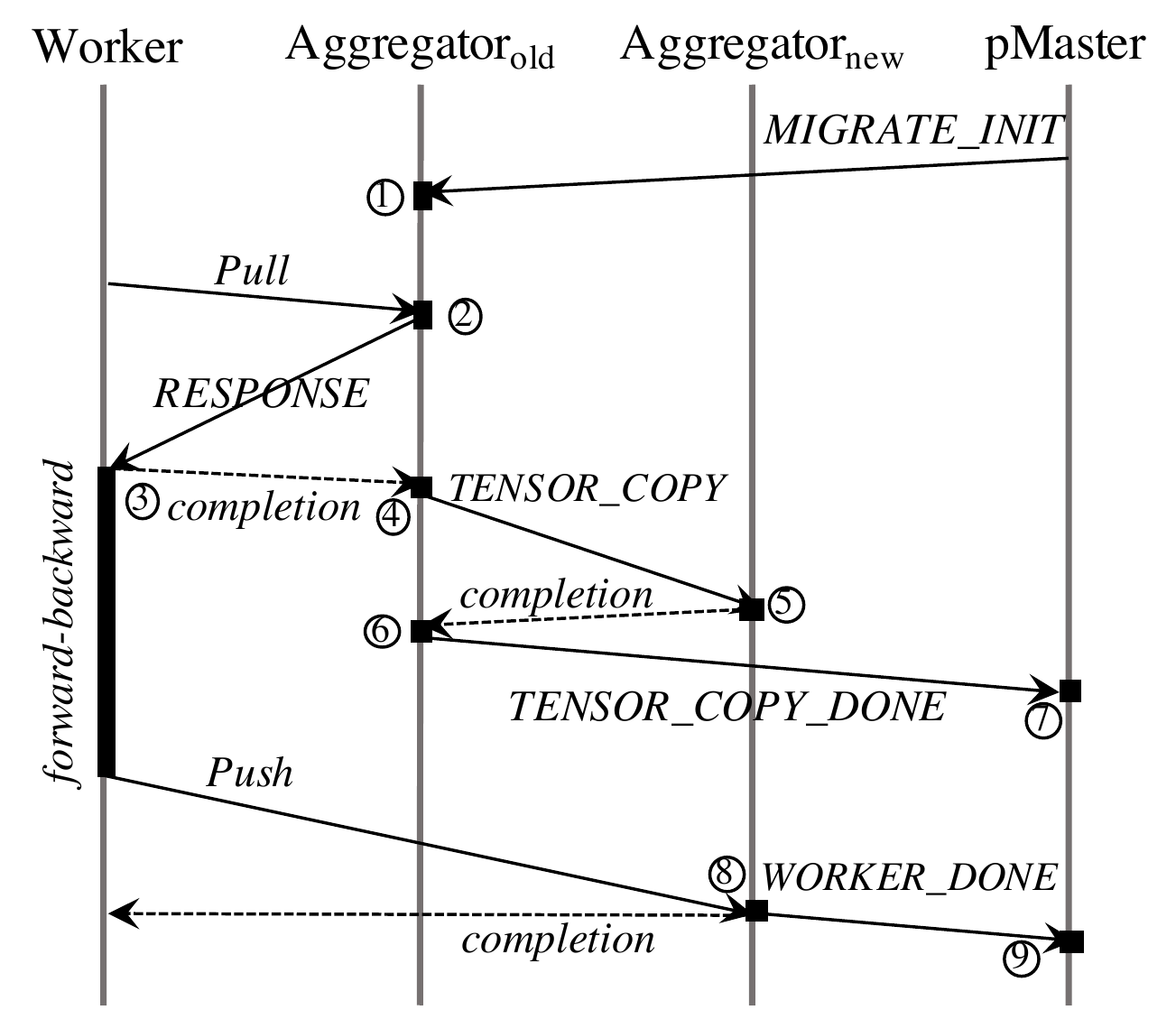}
    \caption{Procedure of tensor migration in {\pservice}.
    $\pserver_{old}$ denotes the old {\pserver}, and $\pserver_{new}$ means the new one.
    \emph{Pull} and \emph{Push} are the original messages in model aggregation.
    The \emph{completion} notifications of data transmission are from the network transport layer.}
    \label{fig:mig-protocol}
\end{figure}

Figure~\ref{fig:mig-protocol} shows the procedure of tensor migration in our design.
{\psmaster} initiates a migration request when a model aggregation needs to be reassigned.
At \textcircled{1}, $\pserver_{old}$ receives the migration request (\emph{MIGRATE\_INIT}) and keeps the information of the tensor and $\pserver_{new}$. 
When the tensor is needed (\emph{Pull}) by the workers at the beginning of next iteration (\textcircled{2}), $\pserver_{old}$ embeds the information of $\pserver_{new}$ into the response and sends to the workers.
Once workers receive the information (\textcircled{3}), their {\psagent}s update the mapping table for the tensor. Then, workers can continue the \emph{forward-backward} computation. 
Meanwhile, $\pserver_{old}$ copies the contents (\eg, metadata, and model parameters) of the tensor (\emph{TENSOR\_COPY}) to $\pserver_{new}$ once the response to workers completes (\textcircled{4}).
At $\pserver_{new}$ side, it adds the incoming tensor into its tensor list (\textcircled{5}). When the copy finishes (\textcircled{6}), $\pserver_{old}$ will notify {\psmaster} (\emph{TENSOR\_COPY\_DONE}) that the tensor data has arrived at $\pserver_{new}$.
After getting the gradient of the tensor in \emph{backward} computation, workers pushes their results to $\pserver_{new}$.
Once those local results arrive (\textcircled{8}), $\pserver_{new}$ also notifies {\psmaster} (\emph{WORKER\_DONE}) that the workers have the updated tensor assignment.
The migration request completes after {\psmaster} receives notifications from both $\pserver_{old}$ and $\pserver_{new}$.

\begin{table}[!t]
  \centering
  \small
  \caption{Time overhead of migrating all tensors in a model.}
  \begin{tabular}{rcccc}
    \toprule
    & \alexnet & \vgg19 & \lm & \bert \\
    \midrule
    Overhead (ms) & $13.6$ & $21.5$ & $40.6$ & $43.8$ \\ 
  \bottomrule
  \end{tabular}
  \label{a_eval:tensor-migration}
\end{table}

\paragraph{Negligible Time Overhead.} 
We measure the time overhead of migrating all tensors in a model from one group of {\pserver}s to the other when the job is running,
and take the averaged value from $10$ runs for each model.
From the view of training jobs, they are only suspended for \emph{tens of milliseconds} 
by the reassignment (Table~\ref{a_eval:tensor-migration}).
The actual durations of those migration operations are much longer, 
most of which are hidden under the computing time at worker side.
Compared with the existing approach of migrating model aggregation (pause, checkpoint, and resume),
which halts the training job for \emph{tens of seconds}~\cite{gandiva, tiresias-nsdi19},
the migration operation in {\autops} has negligible time overhead.
{\autops} uses {protobuf} library to format the data before sending to the network.
It introduces unavoidable overhead (e.g., data copy) by several milliseconds to each reassignment.
The migration operation in {\autops} could be further optimized if it directly applies remote data access feature from {RDMA} networks.

\paragraph{Data Consistency.}
There are two data consistencies that should be guaranteed during tensor migration:
(1) the mapping table among {\psagent}s, and
(2) the master copy of the migrating tensor between $\pserver_{old}$ and $\pserver_{new}$.
{\pservice} merges the procedure of tensor migration into the iterative training procedure.
The information of $\pserver_{new}$ is added into the response message of \emph{Pull} request.
The corresponding mapping table of each {\psagent} can be updated, as long as its worker receives the updated tensor.
Thus, the consistency of the mapping table among {\psagent}s is guaranteed. 
Besides, $\pserver_{new}$ will not execute model update on the tensor that is under migration until the tensor data is completely copied (\emph{TENSOR\_COPY}) from $\pserver_{old}$.
This insures the right version of tensor at $\pserver_{new}$ is used in the following model aggregation.

\section{Problem Definition of Model Aggregation Assignment}
\label{appx:probelm}

\begin{table}[!t]
    \centering
    \small
    \caption{Notations}
    \begin{tabular}{cl}
      \toprule
      Notation & Description \\
      \midrule
        $C_n$ & Execution cycle of {\pserver} $n$\\ 
        $D_j$ & Profiled iteration duration of job $j$ \\
        $d_j$ & Current iteration duration of job $j$\\
        $e_t$ & the execution (CPU) time of task $t$ \\
        $W_n$ & Total execution time of tasks on Aggregation $n$\\
        $L_n$ & Performance (\ie, training speed) loss of job $j$\\
        $\mathbb{N}$ & Set of allocated {\pserver}s\\
        $\mathbb{J}$ & Set of training jobs\\
     \bottomrule
    \end{tabular}
    \label{tab:problem-notion}
\end{table}

Given a set of training jobs ($\mathbb{J}$) and a set of allocated {\pserver}s ($\mathbb{N}$),
how to assign the model aggregation tasks to the limited number of {\pserver}s 
with the minimal performance loss?

The variable in this problem is $p_{tn}$.
It is a binary variable, and indicates whether model aggregation task $t$ is assigned to {\pserver} $n$ or not.

The objective function is to minimize the maximal performance loss among all jobs, which is expressed as:
\begin{equation*}
    \text{minimize} \quad \max_{j \in \mathbb{J}}\{L_j\}
\end{equation*}

There are two constraints:
\begin{align}
    \sum_{n \in \mathbb{N}} p_{tn} = 1, \quad &\forall_{j \in \mathbb{J}, t \in j} \label{ip:c1}\\
    W_n \le C_n, \quad &\forall_{n \in \mathbb{N}} \label{ip:c2}
\end{align}
\begin{align*}
    C_n &=  \max \limits_{j \in \mathbb{J}, t \in j}(p_{tn} \cdot D_j) \\
    d_j &= \max \limits_{t \in j, n \in \mathbb{N}}(\frac{C_n}{\floor*{\frac{C_n}{D_j}}} \cdot p_{tn})\\
    W_n &= \sum_{j \in \mathbb{J}} \sum_{t \in j} (p_{tn} \cdot e_t \cdot \floor*{\frac{C_n}{d_j}}) \\
    L_j &= \frac{d_j - D_j}{d_j}
\end{align*}

The first one means each model aggregation can only be assigned to a single {\pserver}.
The second one requires that {\pserver}s should not be overloaded within each execution cycle.

\section{Mitigating Network Interference}
\label{appx:network}
\begin{figure*}[!t]
    \centering
    \includegraphics[scale=0.53]{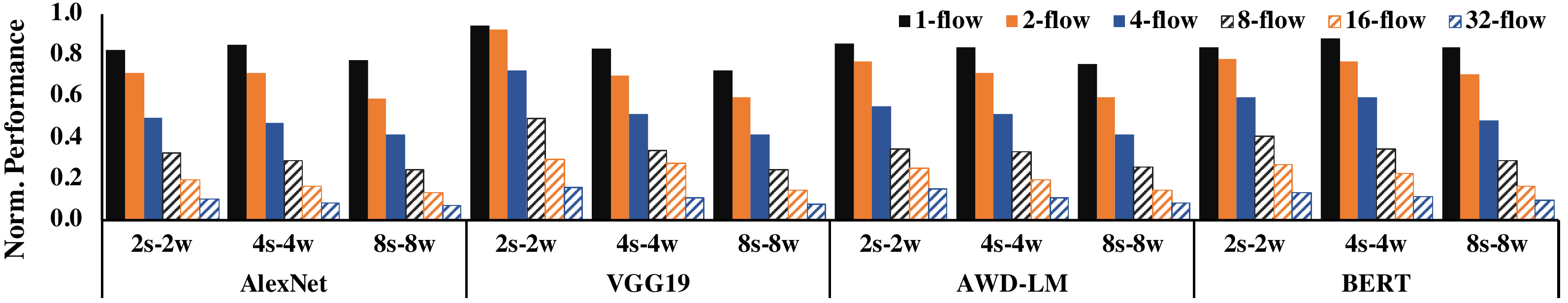}
    \caption{Performance of DDL training jobs when running with network interferences.
    Each model is trained with multiple distributed settings. 
   "2s-2w" means there are 2 PS servers and 2 workers.
    Each server and worker runs on an individual machine.
    In each job, one of its PS servers is interfered by the collocated egress bandwidth flows.
    The performance of each job is normalized to its original performance when running alone.}
    \label{fig:moti-perf-loss-network}
\end{figure*}

\begin{figure*}[!t]
  \centering
  \includegraphics[scale=0.52]{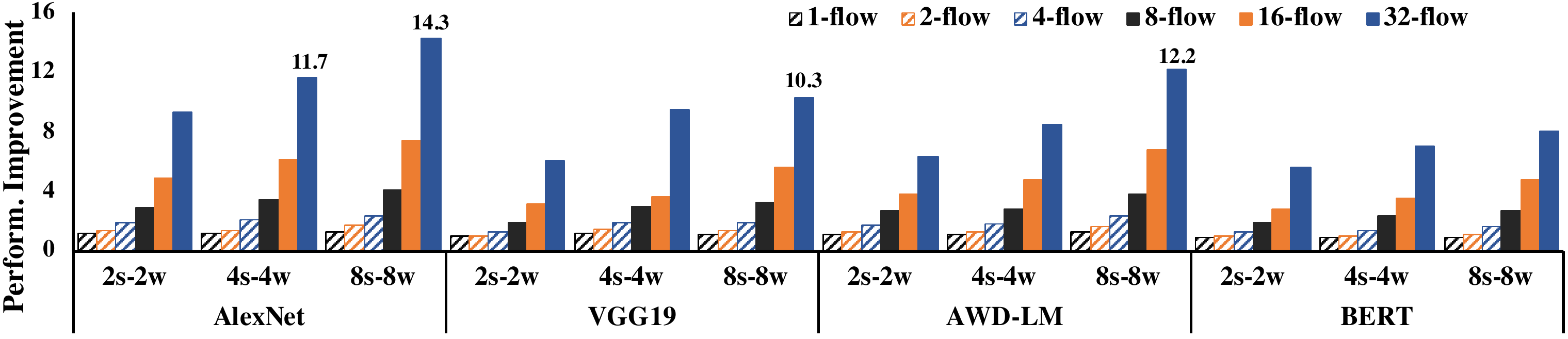}
  \caption{Performance Improvements from {\autops} when there are network interferences.
  No additional {\pserver} is allowed to allocate when {\autops} re-assigns model aggregation workload.
  }
  \label{eval:network-perf-imp}
\end{figure*}

\subsection{Vulnerability to Network Interference}
\label{sec:network-issue}
In current shared clusters, high-speed networks (\eg, RDMA, and DPDK) have been 
deployed to accommodate the increase of computing power (\eg, GPU accelerator)~\cite{atc19-shared-cluster}.
However, due to the dynamic network interferences, the performance benefit from the high-speed network is not always realized for DDL training jobs.

\paragraph{Source of Network Interference.}
First, DL training cannot individually run in the cluster.
There are many other applications~\cite{snorkel-pvldb17} running to prepare the datasets for those training jobs. 
The data-processing frameworks (\eg, Hadoop, and Spark) that those applications rely on have been built on high-speed networks~\cite{rdma-hadoop, rdma-spark} for better performance.
In addition, the cluster itself runs numerous background services to support those 
applications, such as distributed storage~\cite{gpfs-fast02, hyperloop-sigcomm18, storm-systor19}, key-value store~\cite{herd-sigcomm14, fasst-osdi16}, and database systems~\cite{rdma-database-sigmod16}.
All of them share the high-speed network, thus making network availability 
dynamic for DDL training.

The performance (\ie, training speed) of DDL training can be affected when the network becomes congested.
Figure~\ref{fig:moti-perf-loss-network} shows the performance of DDL training jobs when running with network interferences.
We ran experiments on a cluster with $100$ Gbps RDMA network.
Each model is trained on \mxnet, a popular DL framework, with different 
number of workers (w) and PS servers (s).  
In each job, the most heavy-loaded PS server competes for use
of egress bandwidth of the host NIC with other background flows.
Most jobs suffer performance loss in the presence of network interferences,
which becomes worse with the increase of the number of background flows.
For example, the performance loss of \alexnet (2s-2w) rises from $50$\% 
to $90$\% when the number of background flows increases from 4 to 32.
The jobs of \resnet152 are more robust than others. 
This is because most of the tensors in \resnet model are very small; as a result, their network flows are tiny and are not blocked in the network for a long time.

In DDL training, the network load in model aggregation is statically mapped to the physical machines once the job starts.
First, which server instance should handle the aggregation of a tensor is determined 
by the tensor assignment scheme in the DL framework (\eg, \mxnet, and \tensorflow).
However, those schemes (\eg, round-robin) only make static decisions and have not 
yet placed infrastructure information into consideration.
So, the amount of data transferred between each pair of worker and server is constant.
In addition, the server placement also remains fixed once the resource assignment 
is made by the cluster manager. 
This static mapping between the network load and the physical resource makes DDL 
training vulnerable to the network interferences.

\subsection{Experiments}
\label{evl:network}

{\autops} can mitigate the impact of non-transient network interferences through dynamic workload assignment.
Once performance loss and network interferences are detected, it will migrate the model aggregations from the affected {\pserver} to another one that has sufficient remaining resource.

To verify {\autops}'s ability of mitigating network interferences, we replace the parameter server system ({pslite}) with {\autops}, and rerun the experiments in Section~\ref{sec:network-issue}.
The most heavily-loaded {\pserver} is congested by synthetic network background flows (with $1$ MB message size) in the egress port of the host NIC.
Besides, one extra condition is added here. There in no additional {\pserver} available for {\autops} to use, which is designed to see how {\autops} performs without allocating more resource.
Figure~\ref{eval:network-perf-imp} shows the performance improvement from {\autops} when a DDL training job is affected by network interferences.
In general, {\autops} has larger improvements when the network gets more and more congested.
When there are $32$ background flows, {\autops} improves jobs' performance from $5.6\times$ to to $14.3\times$. 

Although, {\autops} uses fewer useable {\pserver}s because of network interferences,
it can still deliver $100\%$ of the job performance in some cases, such as {\vgg19} (4s-4w) and {\lm} (8s-8w).
This means there are free resources among the original {\pserver}s.
As long as the performance of job is higher than the {\lowperf} limit, {\autops} will not re-assign a job even if it detects network interferences.
For example, {\autops} does not migrate the workload of the {\vgg19} (2s-2w) job. 
Because it has $93\%$ of its performance remaining when there are two background flows interfering.

\end{document}